\documentclass[superscriptaddress,prl,twocolumn,amsmath,amssymb,floatfix,showpacs]{revtex4-1}
\usepackage{graphicx}% Include figure files
\pdfoutput=1
\usepackage{bm}% bold mat
\usepackage{amssymb}
\usepackage{color}
\usepackage{epsfig}
\usepackage{subfigure}
\usepackage{hyperref}
\hypersetup{
    pdfstartview={FitH},
   pdfpagemode={None}
   }
\newcommand{\be}{\begin{equation}}
\newcommand{\ee}{\end{equation}}
\begin{document}

\title{Theory of interacting dislocations on cylinders}

\author{Ariel Amir}

\affiliation{Department of Physics, Harvard University, Cambridge, MA 02138, USA}

\author{Jayson Paulose}

\affiliation{Harvard School of Engineering and Applied Sciences, Cambridge, MA
02138, USA}

\author{David R. Nelson}

\affiliation{Department of Physics, Harvard University, Cambridge, MA 02138, USA}
\begin{abstract}
We study the mechanics and statistical physics of dislocations interacting
on cylinders, motivated by the elongation of rod-shaped bacterial
cell walls and cylindrical assemblies of colloidal particles subject
to external stresses. The interaction energy and forces between dislocations
are solved analytically, and analyzed asymptotically. The results
of continuum elastic theory agree well with numerical simulations
on finite lattices even for relatively small systems. Isolated dislocations
on a cylinder act like grain boundaries. With colloidal crystals in
mind, we show that saddle points are created by a Peach-Koehler force
on the dislocations in the circumferential direction, causing dislocation
pairs to unbind. The thermal nucleation rate of dislocation unbinding
is calculated, for an arbitrary mobility tensor and external stress,
including the case of a twist-induced Peach-Koehler force along the
cylinder axis. Surprisingly rich phenomena arise for dislocations
on cylinders, despite their vanishing Gaussian curvature.

\end{abstract}


\maketitle
\section{Introduction} Defects in crystals such as dislocations have
been studied extensively for more than seven decades \cite{orowan,taylor,burgers},
and their importance in condensed matter physics and material science
is widely recognized \cite{hirth}. Systems of dislocations in both
two and three dimensions can be realized experimentally. Mechanical
properties of bulk metals are strongly affected by the dynamics of
dislocation lines within them \cite{hirth}, and two-stage melting
of a two-dimensional crystal can be driven by dislocations \cite{two_d_melting}.
Interesting applications involve a two dimensional particle array
with a periodic boundary conditions in one direction: a cylindrical
crystal. One such example involves interacting colloids on the surface
of a liquid film coating a solid cylinder, where repulsive forces
give rise to the self-organized emergence of a two dimensional crystalline
solid. Defects in colloidal assemblies on the related curved surfaces
of capillary bridges were recently studied experimentally and theoretically
\cite{irvine}. Here, the Gaussian curvature can be positive or negative;
the zero Gaussian curvature of a cylinder is a special case \cite{gillette}.
The growth of the cell walls of rod-shaped bacteria provides a biophysical
example. Their geometry can be approximately described by a cylinder,
and in a recent study \cite{nelson_review,amir_nelson_pnas} we have
argued that one may regard cell wall elongation as mediated by dislocation
climb \cite{hirth}. As a final motivation we note that the hydrodynamic
interactions of vortices on cylinders is mathematically similar to
those of dislocations, albeit simpler since in this case the bare
interactions on a flat surface are isotropic \cite{lamb}. Vortices
on superfluid Helium films were studied in a cylindrical geometry,
in order to model superfluids in porous materials \cite{machta}.
A type II superconducting, hollow wire could have similar interacting
vortices on its cylindrical surface.

\begin{figure}
\includegraphics[width=4cm]{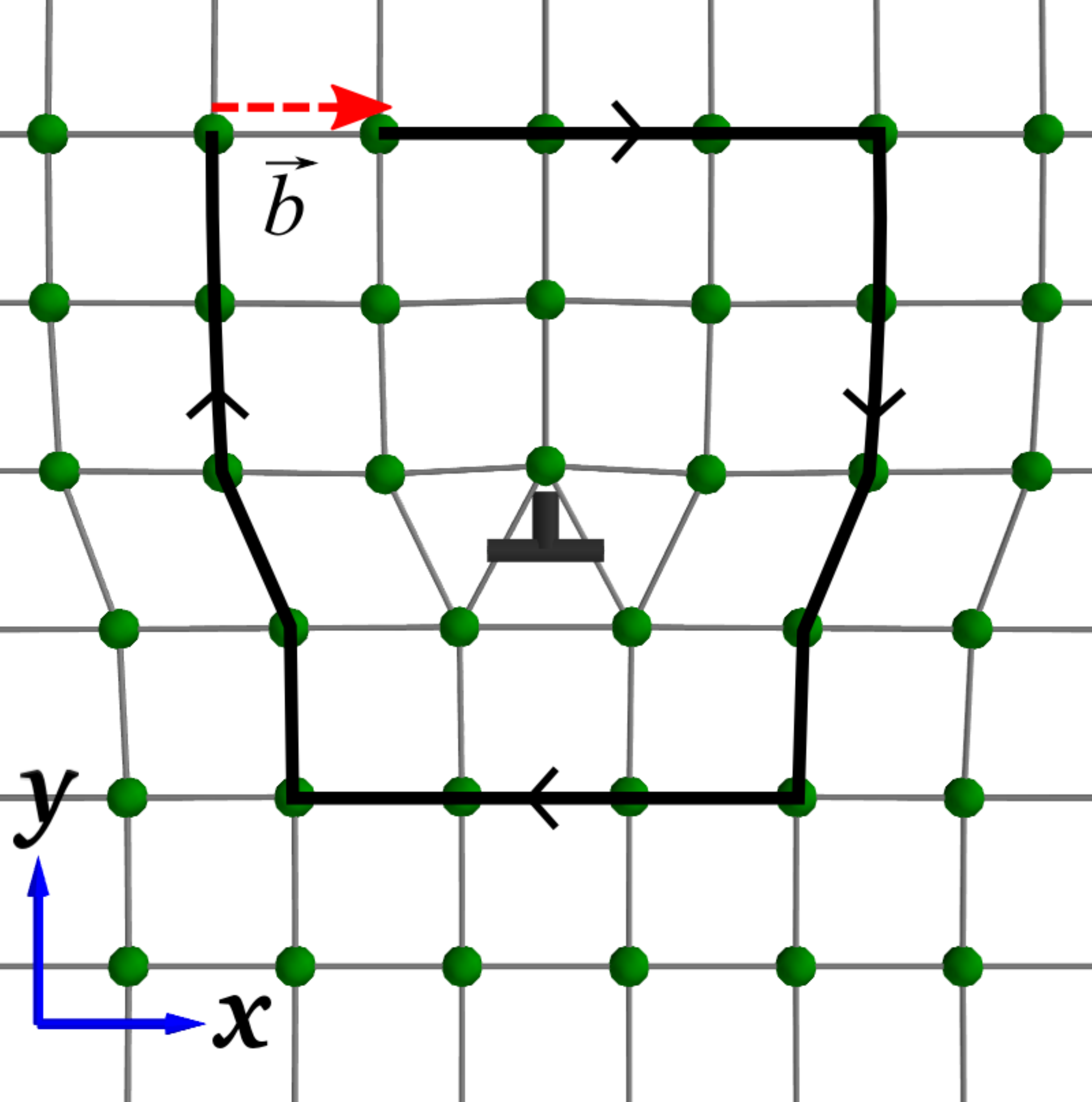} \includegraphics[width=4cm]{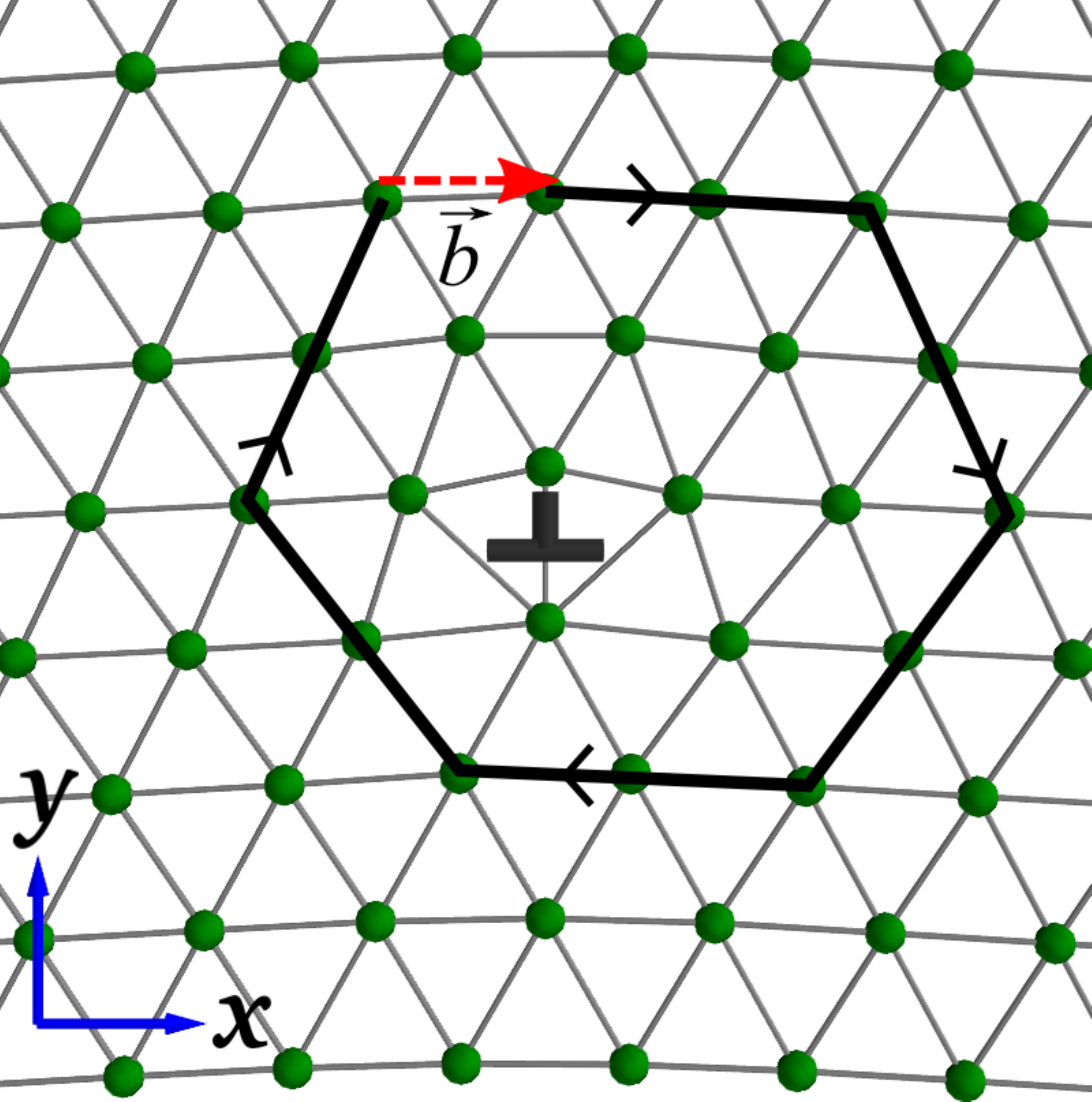}

\caption{Lattice configuration corresponding to a single dislocation with Burgers
vector in the positive $x$ direction for a square (left) and triangular
(right) lattice of masses and harmonic springs relaxed to their equilibrium
configuration. The Burgers vector is defined as the deficit (red arrow)
of a clockwise loop that would be closed in the absence of a dislocation;
hence $\vec{{b}}=b\hat{x}$ in both cases, where $b$ is the lattice
constant. The `T' symbol encodes both the position and orientation
of the dislocation; the leg of the inverted `T' points in the direction
of added material.}

\label{fig_burgersvector}
\end{figure}

Fig. (\ref{fig_burgersvector}) illustrates the simplest dislocation
in two-dimensional square and triangular lattices. For the square
lattice, the dislocation can be thought of as the termination of a
semi-infinite row of lattice points added to the upper half of an
otherwise perfect lattice. For the triangular lattice, the dislocation
represents the termination of \textit{two} semi-infinite rows of particles,
inclined at $\pm30^{\circ}$ to the vertical. Generally, the dislocation
can be characterized by a topological invariant called the Burgers
vector, defined by the Burgers circuit \cite{hirth}, $\vec{{b}}=\oint\frac{\partial u}{\partial l}dl$
(see Fig. \ref{fig_burgersvector}). It is immediately apparent that
such defects distort the crystalline lattice in their vicinity. For
both the triangular and square lattice, it is possible to associated
a particular set of neighbors to each lattice site by assigning near
neighbor bonds. This allows us to define the dislocations using the
number of neighbors: in the square lattice, the dislocation core has
three sites with five nearest neighbors. In the triangular lattice,
the Voronoi construction shows that the dislocation is characterized
by a five-fold site (a site with five nearest neighbors) bonded to
a seven-fold site.

The structure of the manuscript is as follows. We first describe two
scenarios, one motivated by cell wall growth of rod-shaped bacteria
and the other by colloidal crystals on cylinders, and show that they
require understanding the dynamics of interacting dislocations on
a rectangular and triangular lattice, respectively. In contrast to
the predominantly climb dislocation dynamics relevant to elongating
bacteria \cite{amir_nelson_pnas}, the dynamics appropriate to colloidal
assemblies on cylinders is predominantly glide,\textit{ i.e.}, the
motion is parallel as opposed to perpendicular to the Burgers vector
\cite{hirth}. We analyze the form of the interactions between dislocations
with arbitrary Burgers vector and compare the results to numerical
simulations showing that, surprisingly, good agreement with continuum
elastic theory is already achieved for relatively small systems. We
then proceed to exploit a useful connection between isolated dislocations
on a cylinder and the physics of grain boundaries. Finally, we discuss
related `nucleation' problems: Upon the addition of a force in the
circumferential direction driving defects of opposite sign apart,
there will be a finite unbinding rate at non-zero temperature. We
calculate this rate using Langer's generalization of Kramers' theory
\cite{langer,langevin}, and find interesting geometrical effects
associated with the cylindrical geometry. We also discuss the effect
of a twisting stress applied to the ends of a cylinder coated with
colloids; in this case the strain is relaxed by dislocation pairs
separating predominantly along the cylinder axis. The Airy stress
function for a dislocation on a cylinder is calculated in Appendix
A. In Appendix B, we discuss some subtle aspects of the quantization
of stresses and strains on a cylinder, due to periodic boundary conditions.

\subsection {Bacterial cell wall growth}

\label{bacteria}

\begin{figure}
\includegraphics[width=6cm,height=6cm]{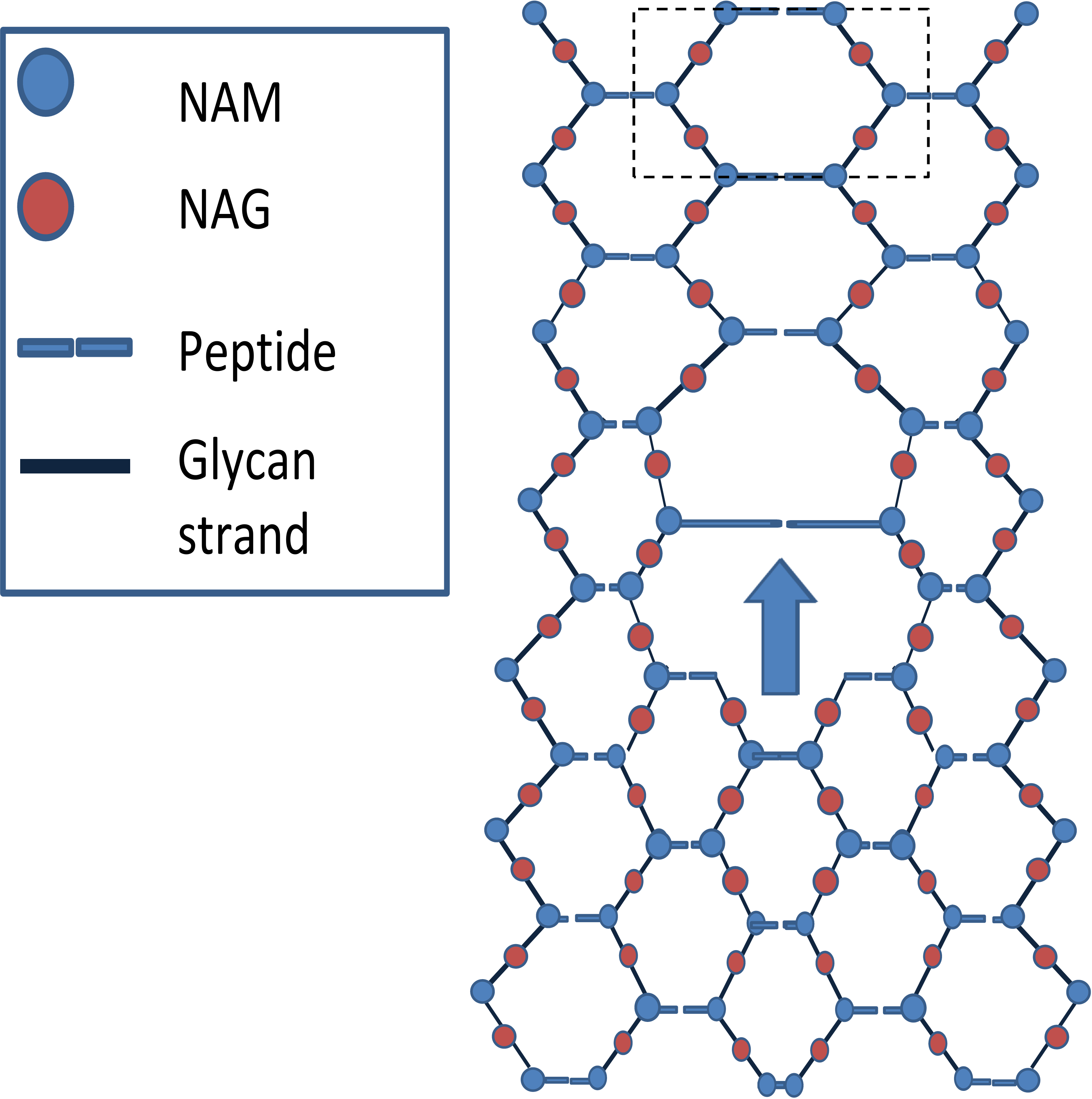}\caption{Schematic illustration of insertions of new glycan strands into the
peptidoglycan mesh of a bacterial cell wall. The new material is inserted
in the vicinity of the blue arrow, which points azimuthally around
the cylinder and sits near the core of the dislocation. The axis of
the cylinder runs horizontally, which is also the direction of the
Burgers vector of this dislocation. Although the the region around
the new insertions is distorted, the connectivity of the structure
is preserved locally due to the simultaneous insertions of \textit{two}
extra glycan strands into the structure, respecting the lattice geometry.
The rectangular (dashed-line) box shows the biologically relevant
unit cell, which is however not the minimal unit cell of the underlying
lattice \cite{nelson_review}.\label{fig:Schematic-illustration}}
\end{figure}

Bacterial cell walls are made of a partly ordered mesh of peptidoglycan
\cite{young}, which can be only a single molecule layer thick in
gram-negative bacteria. While there are still many open questions
regarding architecture and growth, this meshwork is known to consist
of circumferential glycan strands cross-linked by peptides, as shown
schematically in Fig. \ref{fig:Schematic-illustration}. We note that
rod-shaped bacteria with very large aspect ratios can be created by
suppressing the septation process associated with cell division \cite{whitesides}.
To insert new material into the structure, defects in the mesh have
to be created. Inserting a single glycan strand between two existing
ones would not preserve the topology of the network. However, inserting
\textit{two} glycan strands between two existing ones does preserve
the topology. This observation has led to the ``three-for-one''
hypothesis, see for example Ref \cite{pinho}, or, alternatively,
the more symmetric process shown in Fig. \ref{fig:Schematic-illustration}.
Each zig-zag glycan strand is composed of alternating sugar units
called NAM (N-acetylmuramic acid) and NAG (N-acetylglucosamine). For
the unit cell shown in Fig. \ref{fig:Schematic-illustration}, inserting
two glycan strands is equivalent to the addition of a single additional
unit cell. The coarse-grained lattice obtained in this way is rectangular,
and the dislocation can be mediated by edge dislocations climbing
in this lattice in the circumferential direction. The relevant Burgers
vectors point in the direction of the cylinders axis of symmetry \cite{nelson_review,amir_nelson_pnas}.

\begin{figure}
\includegraphics[width=8cm]{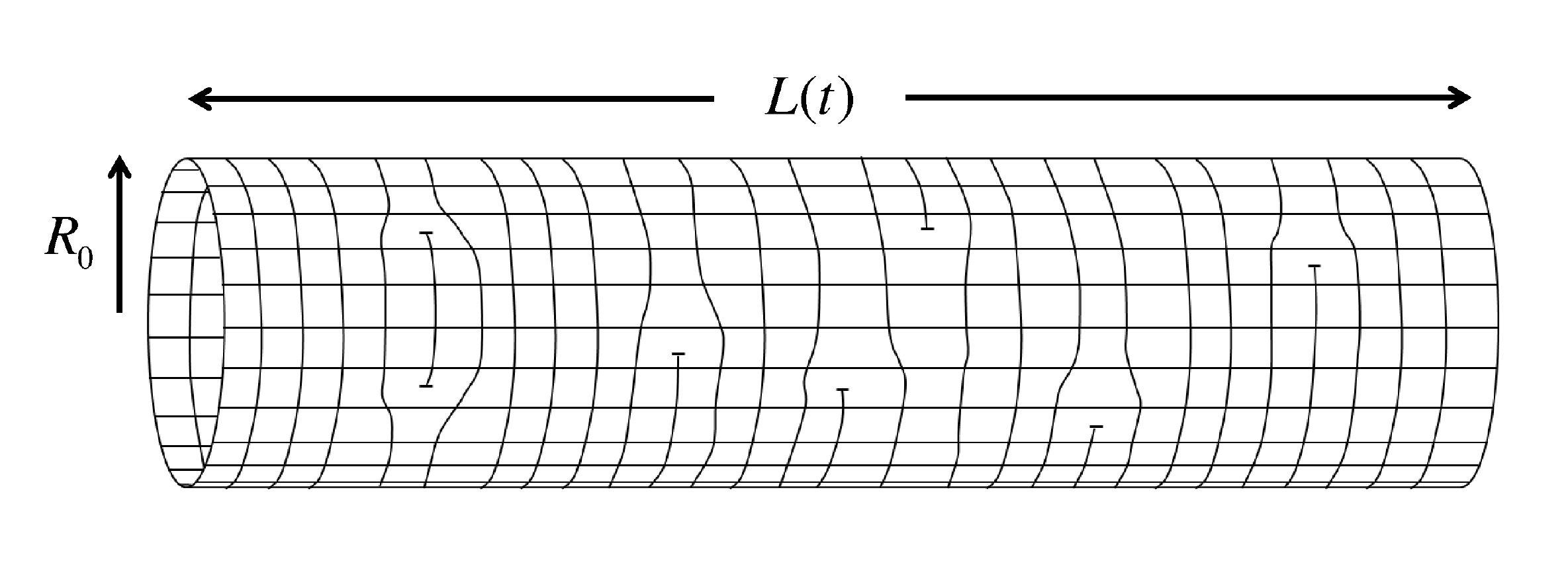}

\caption{Schematic of dislocations (i.e. glycan strand ends) on the cylindrical
portion of a bacterial cell wall of radius $R_{0}$. A subset of these
defects rotate circumferentially when propelled by the addition of
material from inside the bacterium, mediated by strand elongation
machinery on the dislocation cores (not shown). Constant velocity
climb motion of such dislocations leads to exponential elongation
of the cylinder length $L(t)$; see Refs. \cite{nelson_review,amir_nelson_pnas}.}

\label{cylinder}
\end{figure}

Here, we shall simplify the analysis by neglecting the anisotropy
associated with the two-dimensional rectangular lattice, which would
require a non-isotropic elasticity theory with four elastic coefficients
to describe elastic deformations of the structure \cite{ostlund_comment}.
Rather, we shall approximate the systems's free energy by the standard
isotropic form \cite{LL}:

\begin{equation}
F_{elastic}=\int[\frac{\lambda}{2}u_{ii}^{2}(\vec{x})+\mu u_{ij}^{2}(\vec{x})]d^{2}x_{k},\label{eq:expansion}
\end{equation}
where $u_{ij}(\vec{x})=\frac{1}{2}[\partial_{i}u_{j}(\vec{x})+\partial_{j}u_{i}(\vec{x})]$
is the 2d strain tensor. $\lambda$ and $\mu$ are the two-dimensional
\emph{Lamé} \emph{coefficients}. Fig. \ref{cylinder} shows a number
of dislocations in a square lattice. Dislocation climb, mediated by
glycan strand extension machinery, can be affected by interactions
between these dislocations.

\subsection {Colloidal crystals}

Recent experimental advances allow the creation of dislocations in
colloidal assemblies with fascinating interfacial geometries, allowing
study of the interplay of geometry, including Gaussian curvature,
with defects in the lattice structure \cite{bausch,irvine}. In this
case the lattice of colloidal particles is typically triangular, with
lattice vectors that interact only weakly with the directions of principal
curvature. Upon approximating the pair interaction between colloids
as harmonic with spring constant $k_{\text{s}}$ for small displacements
about the equilibrium positions, the lattice can be described elastically
with isotropic effective Lamé coefficients $\lambda=\mu=\sqrt{3}k_{\text{s}}/4$
\cite{seung}. Fig. (\ref{fig_burgersvector}) shows an example of
a single dislocation in a triangular lattice. As we shall show here,
even in the absence of Gaussian curvature, the periodic boundary conditions
associated with a cylindrical geometry give rise to novel phenomena,
not found in an infinite plane (to be referred to as ``flat space''
in the following). In flat space, rotational invariance allows an
arbitrary orientation of the crystallographic axis. On a cylinder,
however, square and triangular lattices can have an energetically
preferred orientation relative to the long axis of the cylinder \cite{kamien_comment}.
With bacterial cell walls, for example, it may be easier to bend the
glycan strands than the alternating amino acid cross-bridges, which
would lead to a preferred orientation on a cylinder. In this paper,
we shall focus primarily on triangular lattices with Bragg rows that
run either azimuthally or along the cylinder axis.

\subsection {Phyllotaxis}

\label{phyllo}

\begin{figure}
\includegraphics[width=8cm]{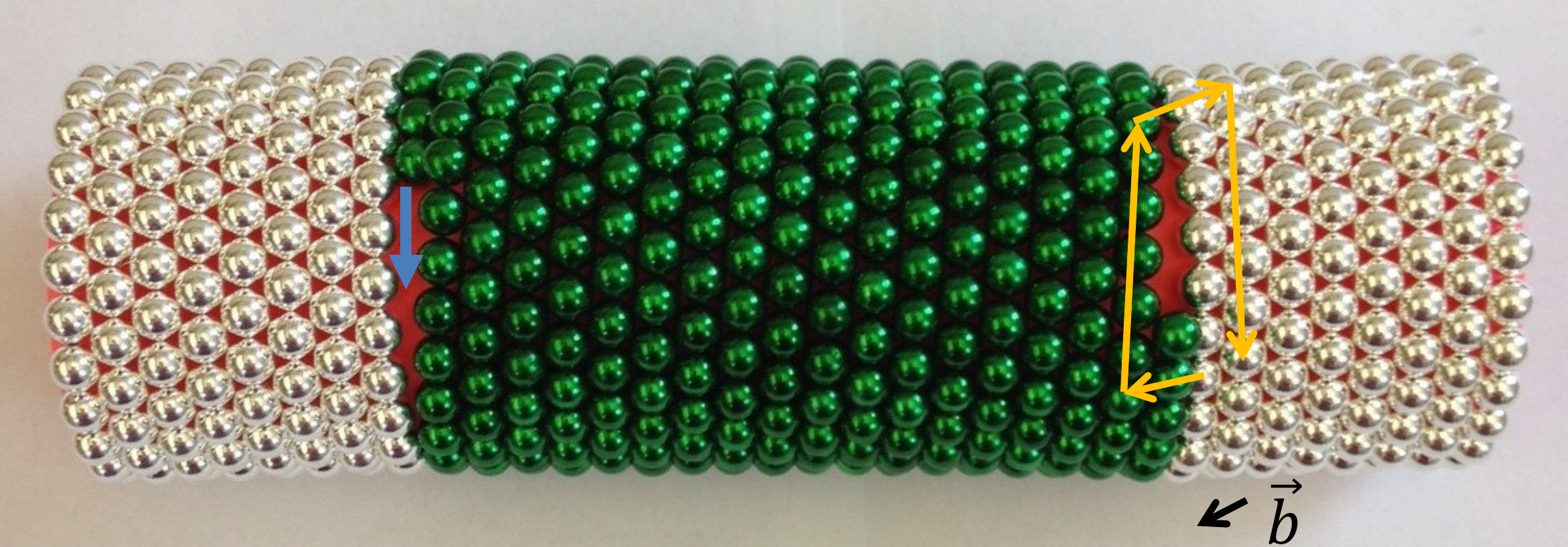}

\caption{Model of a dislocation pair on a cylinder. This structure can systematically
elongate if new particles (green) are added by two counter-rotating
dislocations moving via climb. For one of the dislocations, the climb
direction is indicated by a blue arrow. For the other, the Burgers
circuit around the dislocation is shown (yellow) as well as the resulting
Burgers vector $\vec{b}$. Alternatively, the dislocation pair could
separate by glide (motion parallel to $\vec{b}$) , with motion predominately
along the cylinder, with negligible elongation. In both cases, however,
a triangular lattice with slightly tilted Bragg rows is created in
the center, relative to the purely azimuthal Bragg rows made of silver
particles on the two sides. Although the model was constructed with
magnetic beads interacting via dipole-dipole interactions, we expect
similar configurations for micron-sized colloids assembled on a cylindrical
substrate.}

\label{magnetic}
\end{figure}

In various biological as well as non-living systems, subunits are
arranged in an ordered lattice, wrapped on a cylinder, a particular
case of a phenomenon known as \textit{phyllotaxis} (meaning ``leaf
arrangement'' in ancient Greek). These range from plants \cite{levitov_book,phyllotaxis_review},
rod-shaped viruses and bacterial flagella \cite{erickson,calladine,hartman}
to systems where subunits arrange via magnetic interactions \cite{levitov,crespi1,crespi2}.
Each phyllotactic arrangement can be characterized by two integers
\textit{(M,N)}, such that $Mb\hat{e}_{1}+Nb\hat{e}_{2}=W\hat{y},$
with $\hat{e}_{1}$ and $\hat{e}_{2}$ the two lattice vectors of
the triangular lattice, forming a $60^{\circ}$ angle between them,
and $\hat{y}$ points in the azimuthal direction around the cylinder
of circumference $W=2\pi R$. Such tessellations where recently found
useful also in the context of the Thompson problem on a cylinder,
\textit{i.e.}, how colloids pack in the bulk of a cylinder \cite{mughal}.
In various scenarios the energetically preferred tessellation can
depend on certain external conditions \cite{flagella1,flagella2},
and it is interesting to understand the dynamics of the process through
which this change comes about. It is plausible that the boundary region
between two competing phyllotactic tessellations will consist of one
or several dislocations, see for example Fig \ref{magnetic}; as we
shall show in section \ref{analogy_grain}, a dislocation on a cylinder
is equivalent to a grain boundary. As an example, Ref \cite{hartman}
suggests that such a change in the phyllotactic arrangement of the
tail-sheath of the bacteriophage T4 is driven by 6 dislocations symmetrically
arranged on the circumference of the cylinder. These observations
provide additional motivation for understanding the interactions between
dislocations on a cylinder, which is a necessary step to quantify
the ``dynamical phyllotaxis'' problem sketched above.

\section {Interacting dislocations}

Consider an edge dislocation on the surface of a cylinder, with radius
$R$ and infinite length. We denote the coordinate along the symmetry
axis by $x$, and the other by $y$\emph{,} so that $y$ is periodic
with a period of $W\equiv2\pi R$. For simplicity, and without loss
of generality, let us define the axes' origin at the dislocation core.
In Ref. \cite{amir_nelson_pnas}, some of the components of the stress
tensor produced by edge dislocation with Burgers vector $\vec{{b}}=\pm b\hat{x}$
were evaluated. Here, we shall calculate all components of the stress
produced by dislocations with a Burgers vector $\vec{{b}}$ either
in the $\pm b\hat{x}$ or $\pm b\hat{y}$ direction. For a lattice
of any symmetry or orientation with respect to the cylinder axis we
could always decompose the dislocation's Burgers vector into two orthogonal
components in these directions. Within linear elasticity (and assuming
isotropic elastic constants) one could thus use the results presented
below to find the stress field induced by a dislocation with a Burgers
vector of arbitrary orientation.

In flat space, one may consider a \textit{single} dislocation in an
otherwise perfect crystal, in which case the energy of the system
diverges logarithmically with the system size. However, the energy
of a single dislocation on a long cylinder of length \textit{L}, in
general diverges as the area $WL$, as the subsequent discussion of
long-range strains will make clear. An important special case is a
Burgers vector in the $\hat{x}$ direction, when the divergence is
only logarithmic in \textit{W} and independent of \textit{L}. In the
case of multiple dislocations on a cylinder, the energy can be decomposed
into the divergent terms mentioned above, as well as the interaction
energies. In the following, we focus on these pairwise interactions,
for two generic Burgers vectors. For configurations where the sum
of Burgers vectors vanish (such as cylinders with periodic boundary
conditions along the cylinder axis) the divergences discussed above
will cancel.

As noted in Ref. \cite{amir_nelson_pnas}, the Laplacian on a cylinder
is equivalent to that in an infinite two-dimensional flat space together
with the periodicity requirement. We exploit this idea by first considering
the stresses without the periodicity requirement. Up to a sign depending
on the direction of the Burgers vector $\vec{b}=\pm b\hat{x}$, the
results for the infinite flat space are given by \cite{hirth}:

\begin{equation}
\sigma_{xx}^{flat}=-Aby(3x{}^{2}+y^{2})/r^{4},\label{sigma_xx}
\end{equation}

\begin{equation}
\sigma_{yy}^{flat}=Aby(x^{2}-y^{2})/r^{4},
\end{equation}

and
\begin{equation}
\sigma_{xy}^{flat}=Abx(x^{2}-y^{2})/r^{4},\label{sigma_xy}
\end{equation}
 with $r^{2}=x^{2}+y^{2}$ and $A\equiv\frac{Y}{4\pi}$, where $Y=4\mu(\mu+\lambda)/(2\mu+\lambda)$
is the two-dimensional Young's modulus. Note that the functional form
of the\emph{ }spatial dependence, being determined by geometry, is
the same as for plane stresses around dislocations in three-dimensional
solids. However, the relationship between $A$ and the elastic constants
is different from a three-dimensional isotropic solid \cite{LL}.
For a dislocation with $\vec{{b}}=\pm b\hat{y}$ one has to take $x\rightarrow y$
and $y\rightarrow-x$ on the right hand side of the above equations.

On a cylinder, a dislocation at $(x_{0},y_{0})=(0,0)$ must be duplicated
at intervals of $W$ in the $y$ direction, to respect the boundary
conditions. Therefore the stresses at a point $(x,y)$ generated by
a a dislocation with $\vec{{b}}=b\hat{x}$ at the origin are given
by:

\begin{align}
\sigma_{xx}^{x} & =\sum_{k=-\infty}^{\infty}\frac{-Ab[y+kW][3x^{2}+(y+kW)^{2}]}{[x^{2}+(y+kW){}^{2}]^{2}},\label{sum_1}
\end{align}

\begin{equation}
\sigma_{yy}^{x}=\sum_{k=-\infty}^{\infty}\frac{Ab(y+kW)[x^{2}-(y+kW)^{2}]}{[x^{2}+(y+kW)^{2}]^{2}},\label{sum2}
\end{equation}
and

\begin{equation}
\sigma_{xy}^{x}=\sum_{k=-\infty}^{\infty}\frac{Abx[x^{2}-(y+kW)^{2}]}{[x^{2}+(y+kW)^{2}]^{2}}.\label{sum3}
\end{equation}
 The superscript on the stresses reminds us that these are the stresses
created by a dislocation with $\vec{b}=b\hat{x}.$ Similar sums represent
the stress produced by an edge dislocation with $\vec{{b}}=\pm b\hat{y}$.

The force on another dislocation at $(x,y)$ due to this stress will
then be given by the Peach-Koehler force \cite{peach,hirth,weertman_comment_PRX}:

\begin{equation}
F_{i}=b_{k}\sigma_{jk}\epsilon_{ijz},\label{peach}
\end{equation}
where $\epsilon_{ijz}$ is the Levi-Civita tensor. Explicitly, we
have for $\vec{{b}}=b\hat{x}$:

\begin{equation}
F_{x}=b\sigma_{xy},\label{force_x}
\end{equation}

\begin{equation}
F_{y}=-b\sigma_{xx}.\label{force_y}
\end{equation}

while for $\vec{{b}}=b\hat{y}$:

\begin{equation}
F_{x}=b\sigma_{yy},\label{force_x-1}
\end{equation}

\begin{equation}
F_{y}=-b\sigma_{xy}.\label{force_y-1}
\end{equation}

\subsection {Summing the series}

\label{forces}

The sums of Eqs. (\ref{sum_1}) through (\ref{sum3}) can be evaluated
using the Sommerfeld-Watson transformation \cite{walker_comment}.
To demonstrate this for the first sum, consider the function $g(z)={\rm {cot}(\pi z)}$,
which has only simple poles of residue unity which lie on the $x$
axis at integer values. The sum of Eq. (\ref{sigma_xx}) can then
be written as the complex contour integral:

\begin{equation}
\oint_{C}g(z)f(z)dz,
\end{equation}
 with:

\begin{equation}
f(z)=-Ab\frac{(y/W+z)[3(x/W)^{2}+(y/W+z)^{2}]}{W[(x/W)^{2}+(y/W+z)^{2}]^{2}}.
\end{equation}
 Since $f(z)\sim1/z$ at large distances from the origin, we can deform
the contour so that it captures only the poles of $f(z)$ (note that
the integral on the circle at infinity vanishes even though the decay
is only $\sim1/z$, due to the ${\rm {cot}(\pi z)}$ term). Upon rewriting
the function $f(z)$ as:

\begin{equation}
f(z)=-(Ab/W)\frac{(y/W+z)[3(x/W)^{2}+(y/W+z)^{2}]}{(z+(y+ix)/W)^{2}(z+(y-ix)/W)^{2}},
\end{equation}
 we see that it has two poles of order 2. Summing the residues gives:

\begin{multline}
\sigma_{xx}^{x}=\oint_{C}g(z)f(z)dz=\\
\frac{iAb\pi^{2}x}{2W^{2}}\left({\rm {csc}^{2}(\pi(y-ix)/W)-{\rm {csc}^{2}(\pi(y+ix)/W)}}\right)\\
-\frac{\pi Ab}{2W}\left({\rm {cot}(\pi(y+ix)/W)+{\rm {cot}(\pi(y-ix)/W)}}\right).\label{sigma_xx_cylinder-x}
\end{multline}
 In a similar fashion one obtains the other components of the stress
produced by a dislocation with $\vec{{b}}=b\hat{x}$:

\begin{multline}
\sigma_{yy}^{x}=\\
-\frac{iAb\pi^{2}x}{2W^{2}}\left({\rm {csc}^{2}(\pi(y-ix)/W)-{\rm {csc}^{2}(\pi(y+ix)/W)}}\right)\\
-\frac{\pi Ab}{2W}\left({\rm {cot}(\pi(y+ix)/W)+{\rm {cot}(\pi(y-ix)/W)}}\right),\label{sigma_yy_cylinder-x}
\end{multline}

\begin{equation}
\sigma_{xy}^{x}=-\frac{Ab\pi^{2}x}{2W^{2}}\left({\rm {csc}^{2}(\pi(y-ix)/W)+{\rm {csc}^{2}(\pi(y+ix)/W)}}\right).\label{sigma_xy_cylinder-x}
\end{equation}

To find the components of the stress tensor due to a dislocation with
$\vec{{b}}=\hat{by}$, we use the previously mentioned substitution
$x\rightarrow y$, $y\rightarrow-x$, which leads immediately to:
\begin{align}
\sigma_{xx}^{y} & =\sigma_{xy}^{x},\label{sigma_xx_cylinder-y}\\
\sigma_{xy}^{y} & =\sigma_{yy}^{x}.\label{sigma_xy_cylinder-y-1}\\
\nonumber
\end{align}
Application of the Somerfeld-Watson transformation to the remaining
component of the stress tensor leads to:

\begin{align}
\sigma_{yy}^{y} & =-\frac{Ab}{2W}\pi[-2\text{coth}[\pi(x-iy)/W]-2\text{coth}[\pi(x+iy)/W]\nonumber \\
 & +\pi\frac{x}{W}\left(\text{csch}[\pi(x-iy)/W]^{2}+\text{csch}[\pi(x+iy)/W]^{2}\right)].\label{eq:sigma_yy_cylinder-y}
\end{align}

\subsection{Asymptotic forms}

It is natural to consider various limits for the stresses. For distances
small compared to the cylinder's radius, it can be checked that the
previous expressions all reduce to the flat space results, as must
be the case. However, for large separations along the $x$ direction
(i.e., the cylinder axis), the behavior is different. For a dislocation
with $\vec{b}=b\hat{x}$, we find that for $|x|\gg W=2\pi R$:

\begin{equation}
\sigma_{xx}^{x}\approx-4\pi^{2}Abe^{-2\pi|x|/W}(|x|/W^{2}){\rm {sin}(2\pi y/W),}\label{sigmax_xx_asympt}
\end{equation}
\begin{equation}
\sigma_{yy}^{x}\approx4\pi^{2}Abe^{-2\pi|x|/W}(|x|/W^{2}){\rm {sin}(2\pi y/W),}\label{sigmax_yy_asympt}
\end{equation}
\begin{equation}
\sigma_{xy}^{x}\approx4\pi^{2}Abe^{-2\pi|x|/W}(|x|/W^{2}){\rm {cos}(2\pi y/W),}\label{sigmax_xy_asympt}
\end{equation}

while for a dislocation with $\vec{b}=b\hat{y}$, we have:

\begin{equation}
\sigma_{xx}^{y}\approx4\pi^{2}Abe^{-2\pi|x|/W}(|x|/W^{2}){\rm {cos}(2\pi y/W),}\label{sigmay_xx_asympt}
\end{equation}
\begin{align}
\sigma_{yy}^{y} & \approx2\pi A(b/W)\cdot Sgn(x)\nonumber \\
 & -4\pi^{2}Abe^{-2\pi|x|/W}(|x|/W^{2}){\rm {cos}(2\pi y/W)},\label{sigmay_yy_asympt}
\end{align}
\begin{equation}
\sigma_{xy}^{y}\approx4\pi^{2}Abe^{-2\pi|x|/W}(|x|/W^{2}){\rm {sin}(2\pi y/W),}\label{sigmay_xy_asympt}
\end{equation}
where $Sgn(x)=x/|x|.$ Thus, all components of the stress tensor decay
exponentially, except for the circumferential stress induced by a
dislocation with $\vec{b}=b\hat{y}$, which approaches to a constant
value exponentially fast. This constant reflects the half-line of
extra material that is inserted throughout the long axis of the cylinder,
leading to a long-ranged stress field, as evident in the first term
of Eq. (\ref{sigmay_yy_asympt}).

\subsection{Energy considerations}

In this section we convert our results for the stresses to the elastic
interaction energies for two dislocations on a cylinder, obtained
by integrating the Peach-Koehler force of Eq. (\ref{peach}). As discussed
above, a generic Burgers vector can be decomposed into $\hat{x}$
and $\hat{y}$ components, and thus we consider three distinct scenarios:

(a) Both dislocations have Burgers vectors in the $\pm\hat{x}$ directions.

(b) One dislocation with $\vec{b}=\pm b\hat{x}$ with another with
$\vec{b}=\pm b\hat{y}$.

(c) Both dislocations have Burgers vectors in the $\pm\hat{y}$ directions.

In flat space, the interaction energy of two edge dislocations with
vectors $\vec{b}_{1}$ and $\vec{b}_{2}$, and with a relative separation
of $\vec{r}=(x,y),$ $r\gg b,$ is given by~\cite{two_d_melting}:

\begin{equation}
E(x,y)=-A\left((\vec{b_{1}}\cdot\vec{b_{2}})\log[\frac{r}{b}]-\frac{(\vec{b_{1}}\cdot\vec{r})(\vec{b_{2}}\cdot\vec{r})}{r^{2}}\right)+2E_{c},\label{energy_flat}
\end{equation}
where $b$ is the lattice spacing, and the effect of higher order
terms in the gradient expansion of Eq. (\ref{eq:expansion}) is given
by the core energy term $2E_{c}$ \cite{fisher_comment}. Note that
unless the Burgers vectors are equal and opposite, the total energy
of the system will also include terms that diverge with the system
size, as previously discussed.

The derivatives of the interaction energy with respect to the coordinates
yield the forces: for example, differentiating Eq. (\ref{energy_flat})
with respect to $x$ or $y$ and using Eqs. (\ref{force_x}) and (\ref{force_y})
gives Eqs. (\ref{sigma_xy}) and (\ref{sigma_xx}). Up to a constant,
we can obtain this interaction energy by integration of the Peach-Koehler
force.

Upon generalizing to the case of the cylinder, where we have already
found an explicit formula for the forces, we can use it to derive
the expression for the interaction energy for case (a) above. Integrating
the force in the $x$ direction, $F_{x}=-b\sigma_{xy}^{x},$ with
respect to $x$ leads to $E(x,y)=Y(y)+C(x,y)$ with:

\begin{multline}
C(x,y)=\frac{Ab^{2}}{2}{\rm {log}}[{\rm sinh}(\pi(x-iy)/W)]+{\rm }\\
+\frac{Ab^{2}}{2}i\pi(x/W){\rm {csc}(\pi y/W){\rm {sinh}(\pi x/W){csch}(\pi(x-iy)/W)}}\\
+C.C.,\label{energy_}
\end{multline}

with $W=2\pi R$. The derivative of $C(x,y)$ with respect to $y$
can be shown to be equal to $F_{y}=b\sigma_{xx}^{x}$, implying that
$Y(y)=const$. The constant can be found be demanding that the expression
reduces to that of flat space for $W\gg x,y$, see Eq. (\ref{energy_flat}).
Our final result for antiparallel Burgers vectors along the cylinder
axis is thus:

\begin{multline}
E_{\hat{x},-\hat{x}}(x,y)=\frac{Ab^{2}}{2}{\rm {log}}[{\rm \frac{W}{\pi b}sinh}(\pi(x-iy)/W)]+{\rm }\\
+\frac{Ab^{2}}{2}i\pi(x/W){\rm {csc}(\pi y/W){\rm {sinh}(\pi x/W){csch}(\pi(x-iy)/W)}}\\
+C.C.\label{energy}
\end{multline}
We have suppressed, for simplicity, the large distance core energy
contribution displayed in Eq. (\ref{energy_flat}). The notation $E_{\hat{x},-\hat{x}}$
denotes that this is the interaction energy of two antiparallel dislocations
with Burgers vectors in the $\pm\hat{x}$ directions. For $x\gg R$,
we find that:

\begin{equation}
E_{\hat{x},-\hat{x}}\approx Ab^{2}\log[\frac{W}{2\pi b}]-2Ab^{2}\frac{|x|}{W}e^{-2\pi|x|/W}\cos(2\pi y/W).\label{energy_xx}
\end{equation}

\begin{figure}[!]
\includegraphics[width=4cm]{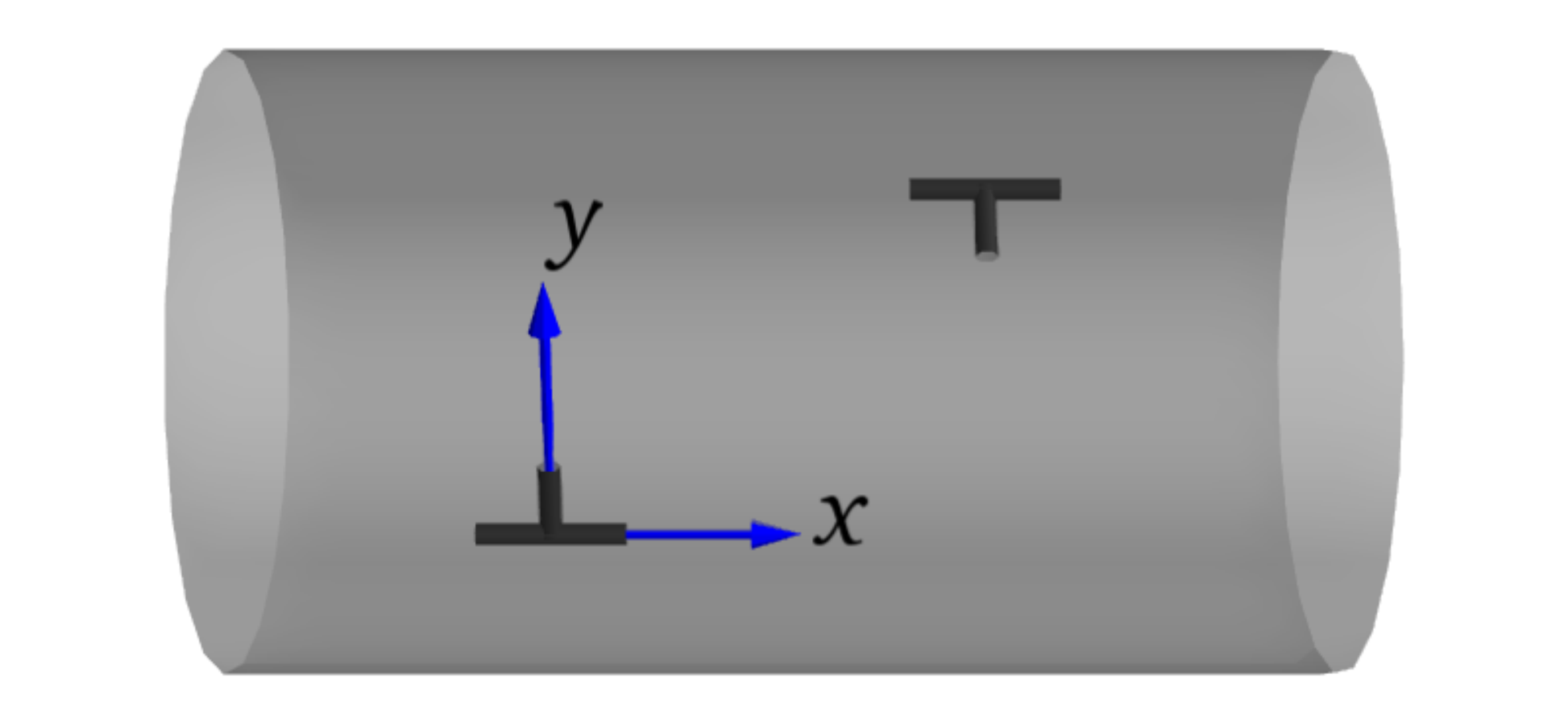}

\includegraphics[width=8cm]{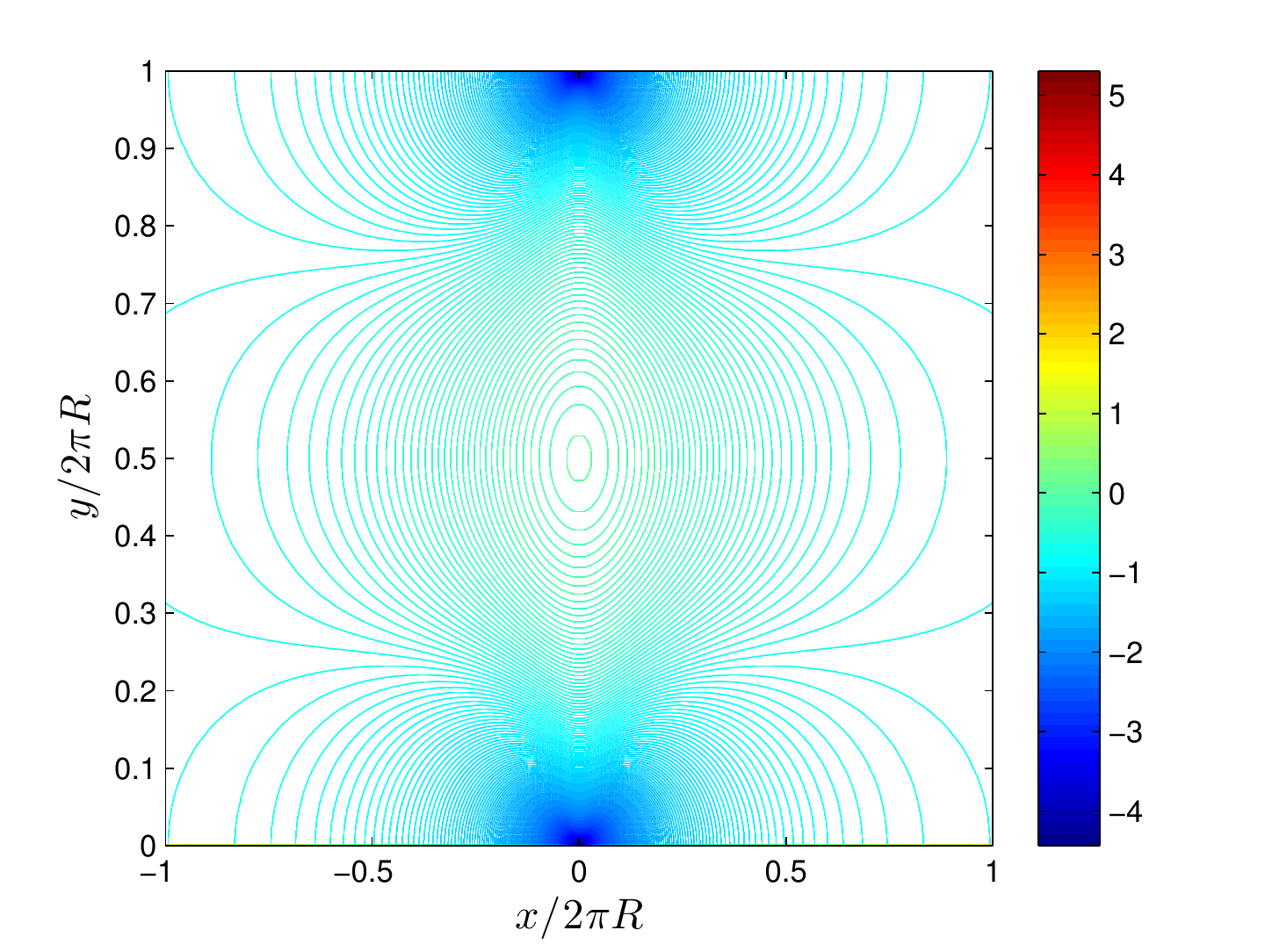} \caption{Equipotential contours for a dislocation on a cylinder at position
$(x,y)$ with $\vec{b}=-b\hat{x}$, interacting with another dislocation
at the origin with $\vec{b}=b\hat{x}$. Energy is measured in units
of $Ab^{2}.$ Note the much lower energies than in Fig. \ref{energy_yy fig}
where the Burgers vector are rotated by $90^{\circ}$.}

\label{contour_plot}
\end{figure}

\begin{figure}[!]
\includegraphics[width=8cm]{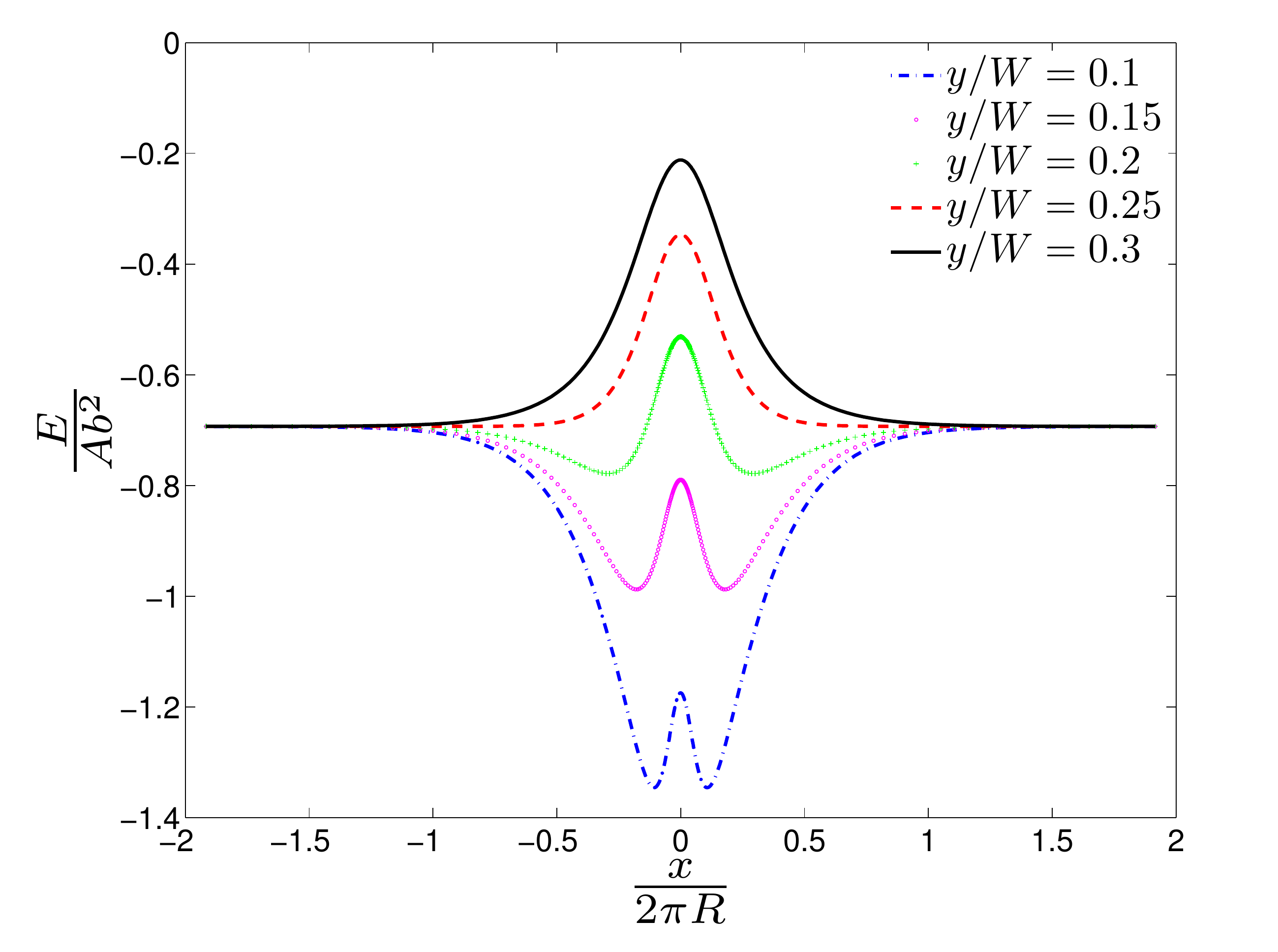} \caption{Slice along the energy contours of Fig. \ref{contour_plot} showing
interaction energy as a function of $x$, for several ratios of $y/W$.
Note the double minima, which become a single maximum for $y\geq W/4$.}

\label{cut_1}
\end{figure}

Fig. \ref{contour_plot} shows the equal energy contours of this interaction
energy. Close to the origin, a cut parallel to the $\hat{x}$-axis
would give a graph with two minima, corresponding to the two dislocations
with separation vector at a 45 degrees angle to the $\hat{x}$-axis
-- the double minima are expected, since close to the origin we are
not sensitive to the finite radius of the cylinder, and this is indeed
the stable configuration of two dislocations in flat space, when climb
processes (motion perpendicular to the Burgers vector) are prohibited
\cite{bruinsma}. The double minima structure at a fixed offset $y=W/10$
is shown in Fig. \ref{cut_1}. However, at larger vertical separations
the minima become shallower and shallower, until at a separation of
$W/4$ along the circumference, there is only a single maximum at
$x=0$. For fixed $y$, the minima obey $\frac{\partial E_{\hat{x},-\hat{x}}(x,y)}{\partial x}\propto\sigma_{xy}^{x}=0$.
Upon equating Eq. (\ref{sigma_xy_cylinder-x}) to zero, we find (aside
from the solution $x=0$):
\begin{equation}
{\rm {tan}(\pi y/W)=\pm{\rm {tanh}(\pi x/W),}}\label{x-vanish}
\end{equation}
 which indeed has a solution at nonzero $x$ provided $y<W/4$.

Taking the $y\rightarrow0$ limit of Eq. (\ref{energy}) (which is
ill-defined for $y=0)$ leads to:

\begin{flalign}
E_{\hat{x},-\hat{x}}(x,y & \to0)=\nonumber \\
 & Ab^{2}\left({\rm {log}[{\rm \frac{2R}{b}{sinh}(\pi x/W)]-\frac{\pi x}{W}{\rm {coth}(\pi x/W)}}}\right).\label{eq:limit_bruinsma}
\end{flalign}
 This formula is equivalent to Eq. (2.1b) in Ref. \cite{bruinsma},
describing the interaction energy of two antiparallel grain boundaries
in the infinite plane. The connection between these two systems will
be elucidated in section \ref{analogy_grain}.

\begin{figure}[!]
\includegraphics[width=4cm]{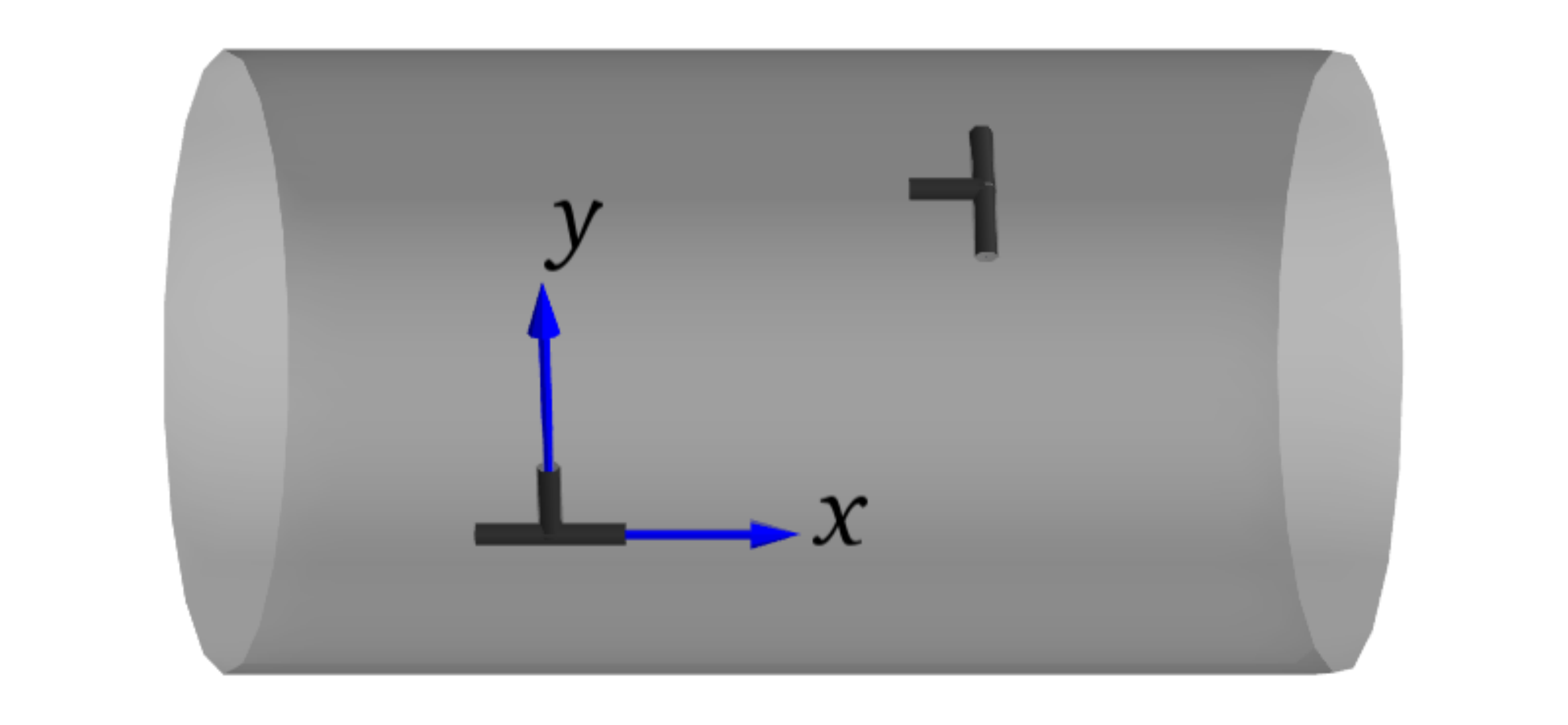}

\includegraphics[width=8cm]{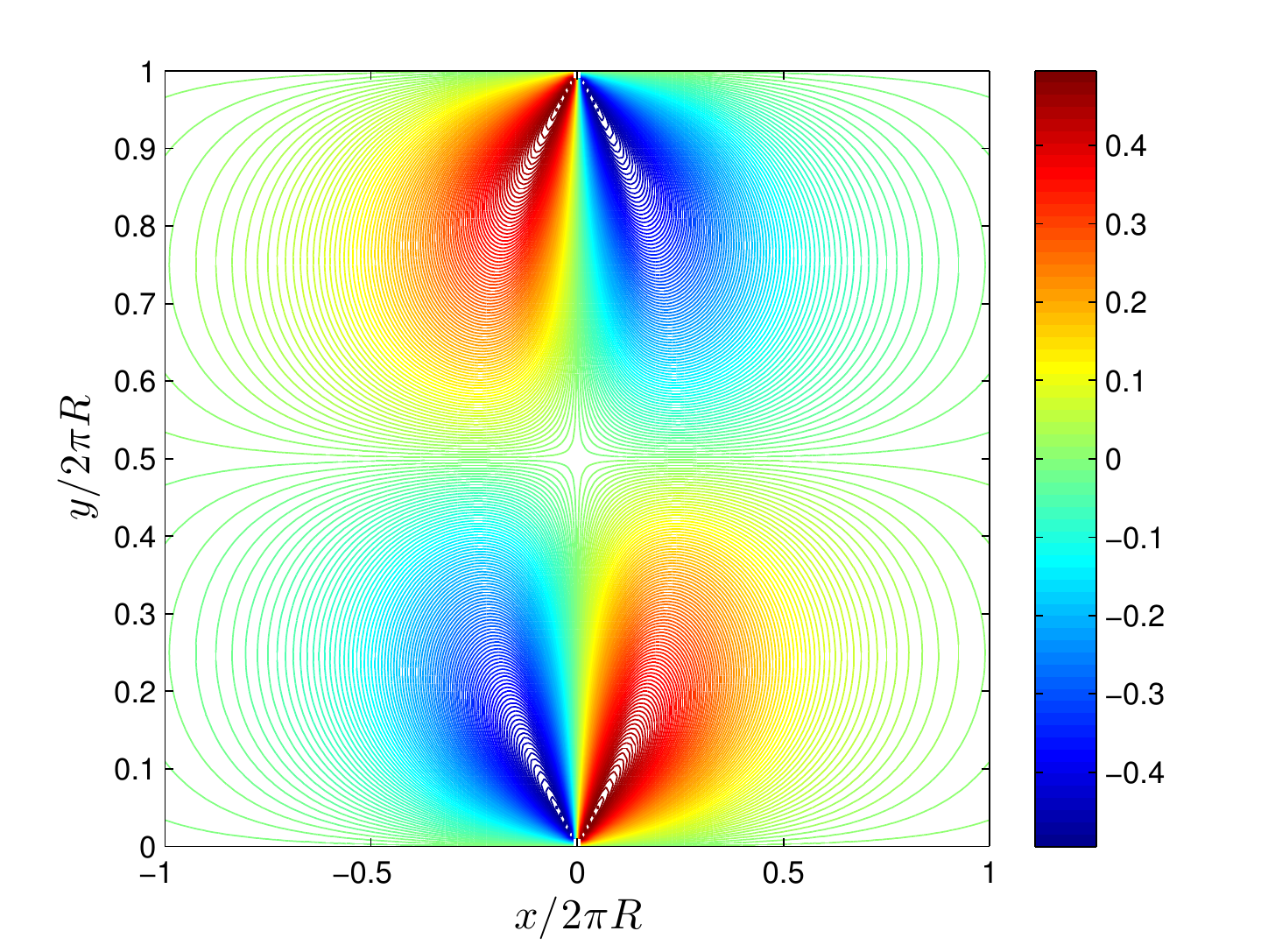} \caption{Equipotential contours for a dislocation at $(x,y)$ on a cylinder
with $\vec{b}=b\hat{y}$, interacting with another dislocation at
the origin with $\vec{b}=b\hat{x}$. Energy is measured in units of
$Ab^{2}.$ Note that the large distance core energy contribution,
analogous to that appearing in Eq. (\ref{energy_flat}), is undefined
without additional dislocations on the cylinder, since the sum of
the Burgers vectors should vanish. Nevertheless, this expression is
useful for charge-neutral dislocation configurations on a cylinder,
whose interactions can be decomposed into pairwise interactions involving
$E_{\hat{x},-\hat{x}}$, $E_{\hat{y},-\hat{y}}$ and $E_{\hat{x},\hat{y}}$;
see Eqs. (\ref{energy_xx}),(\ref{energy-yy}) and (\ref{energy_xy}).}

\label{energy_xy fig}
\end{figure}

In a similar fashion we can find the interaction energy for case (b)
above, of an edge dislocation with $\vec{{b}}=b\hat{x}$ with another
with $\vec{{b}}=b\hat{y}$. The forces on the latter are given by
$F_{x}=\sigma_{yy}^{x}$, $F_{y}=-\sigma_{xy}^{x}$. By integrating
the stresses we find:

\begin{equation}
E_{\hat{x},\hat{y}}(x,y)=-\frac{Ab^{2}\pi\frac{x}{W}\text{sin}[2\pi y/W]}{\text{cos}[2\pi y/W]-\text{cosh}[2\pi x/W]}.\label{energy_xy}
\end{equation}
The equal energy contours of this interaction energy are shown in
Fig. \ref{energy_xy fig}. For $x\gg R$ the expression reduces to:

\[
E_{\hat{x},\hat{y}}\approx2\pi Ab^{2}\frac{x}{W}\text{sin}[2\pi y/W]e^{-2\pi|x|/W}.
\]
For $y=0$, we find that $E_{\hat{x},\hat{y}}=0$. For $x,y\ll W$,
Eq. (\ref{energy_xy}) reduces to the flat expression:
\begin{equation}
E_{\hat{x},\hat{y}}\approx\frac{Ab^{2}xy}{x^{2}+y^{2}}=\frac{Ab^{2}}{\frac{x}{y}+\frac{y}{x}}.
\end{equation}

Since $|c+\frac{1}{c}|\geq2$ for any value of $c$, there is clearly
no divergence of the energy as the two dislocations come together,
with a minimum energy of $-\frac{Ab^{2}}{2}$ obtained for $x=-y.$
The finiteness of the energy as the dislocations as $(x,y)\rightarrow(0,0)$
reflects the vanishing of the logarithmic divergence in Eq. (\ref{energy_flat})
when $\vec{b_{1}}$ and $\vec{b_{2}}$ are perpendicular. As mentioned
before, a dislocation configuration where the sum of Burgers vectors
does not vanish leads to additional terms in the expression of the
total energy diverging with the system size. The above equations only
reflect the pairwise interaction terms.

\begin{figure}[!]
\includegraphics[width=4cm]{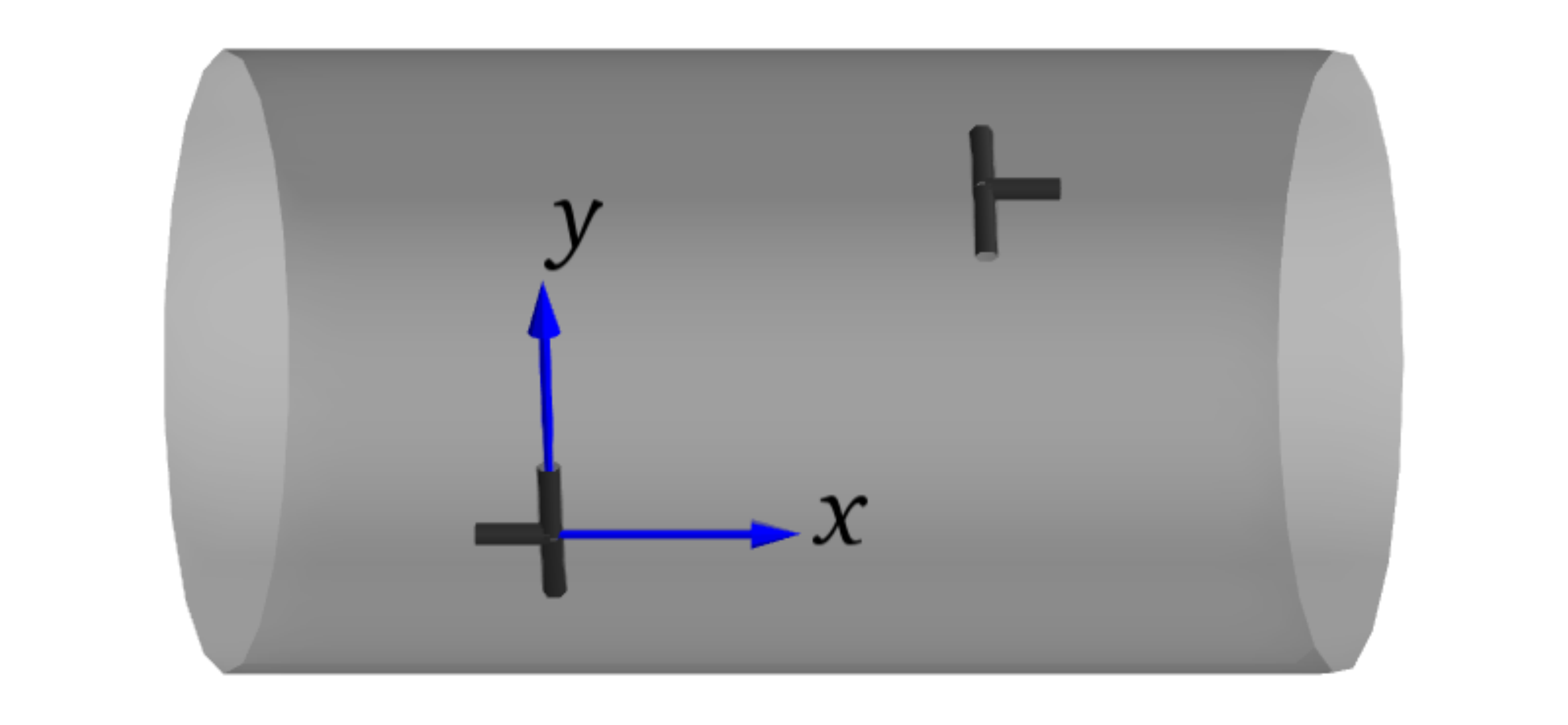}

\includegraphics[width=8cm]{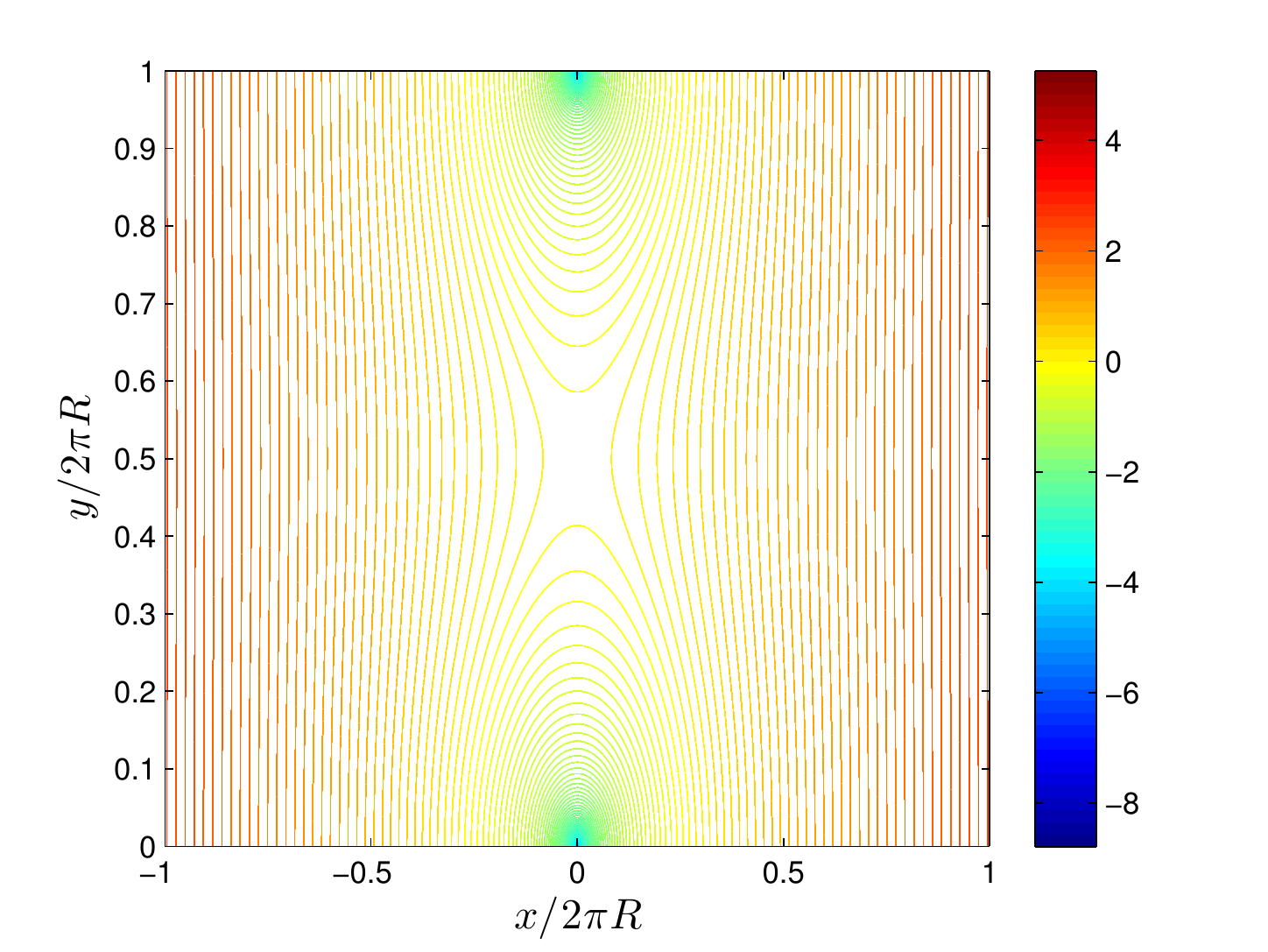} \caption{Equipotential contours for a dislocation at $(x,y)$ on a cylinder
with $\vec{b}=-b\hat{y}$, interacting with another dislocation at
the origin with $\vec{b}=b\hat{y}$. Energy is measured in units of
$Ab^{2}.$ Note that for $x\gg R=W/2\pi$ the energy contours become
parallel to the $y$ axis; so that the forces are primarily in the
$x$ direction. This trend, and the associated linear potential, can
be seen from Eqs. (\ref{force_x-1}), (\ref{force_y-1}),(\ref{sigmay_yy_asympt})
and (\ref{sigmay_xy_asympt}). }

\label{energy_yy fig}
\end{figure}

Finally, in case (c), the force on an edge dislocation with $\vec{{b}}=-b\hat{y}$
induced by another dislocation with $\vec{{b}}=b\hat{y}$ is given
by $F_{x}=-\sigma_{yy}^{y}$, $F_{y}=\sigma_{xy}^{y}$. Upon integration
we find
\begin{multline}
E_{\hat{y},-\hat{y}}(x,y)=\frac{Ab^{2}}{2}\text{log}[\text{\ensuremath{\frac{W}{e\pi b}}sinh}[\pi(x-iy)/W]]\\
-\frac{Ab^{2}}{2}i\pi(x/W)\text{csc}[\pi y/W]\text{\text{sinh}[\ensuremath{\pi}x/W]csch}[\pi(x-iy)/W]+\\
C.C.,\label{energy-yy}
\end{multline}
where $e$ is the base of the natural logarithm. The equal energy
contours are shown in Fig. \ref{energy_yy fig}. The analytic form
of Eq. (\ref{energy-yy}) is similar (but not identical!) to that
for $E_{\hat{x},-\hat{x}},$ see Eq. (\ref{energy}). As before, we
can check that for $x,y\ll R$ this reduces to Eq. (\ref{energy_flat})
(this correspondence requires the extra factor $2/e$ in the logarithm).
For $x\gg R$ we now find, in contrast to Eq. (\ref{energy_xx}),
a linear potential,

\begin{equation}
E_{\hat{y},-\hat{y}}(x,y)\approx\frac{2\pi Ab^{2}}{W}|x|.\label{eq:yy_interaction_large_x}
\end{equation}

In the limit $y\rightarrow0$, similarly to the case of $E_{\hat{x},-\hat{x}}(x,0)$,
we find that:

\begin{equation}
E_{\hat{y},-\hat{y}}(x,0)=Ab^{2}{\rm \left({log}[{\rm \frac{W}{e\pi b}{sinh}(\pi x/W)]+\frac{\pi x}{W}{\rm {coth}(\pi x/W)}}\right).}
\end{equation}

`\subsection{Stresses and energetics for Burgers vectors at arbitrary inclination angle}

A Burgers vector $\vec{b}$ that makes an angle $\theta$ with the
$x$ axis, can always be decomposed into Cartesian components parallel
and perpendicular to the cylinder axis. Since we assume \emph{linear}
elasticity, the relevant stress fields follow from the superposition
of the solutions obtained previously:
\begin{equation}
\sigma_{ij}^{\theta}=\sigma_{ij}^{x}\cos\theta+\sigma_{ij}^{y}\sin\theta,\label{sigma_ij_theta}
\end{equation}
where $\sigma_{ij}^{x}$ and $\sigma_{ij}^{y}$ appear in Eqs. (\ref{sigma_xx_cylinder-x}--\ref{eq:sigma_yy_cylinder-y}).

We now determine the interaction energy of this dislocation with another
dislocation, whose Burgers vector forms an angle $\alpha$ with the
$x$ axis (on a triangular lattice, if the Burgers vectors have their
minimum allowed lengths, the difference of the two angles $\theta$
and $\alpha$ will be a multiple of $\pi/3$). The force on this dislocation
is then:
\begin{multline}
F_{x}=b[\sigma_{xy}^{x}\cos\theta\cos\alpha+\sigma_{xy}^{y}\sin\theta\cos\alpha\\
+\sigma_{yy}^{x}\cos\theta\sin\alpha+\sigma_{yy}^{y}\sin\theta\sin\alpha].
\end{multline}

\begin{multline}
F_{y}=b[-\sigma_{xx}^{x}\cos\theta\cos\alpha-\sigma_{xx}^{y}\sin\theta\cos\alpha\\
-\sigma_{xy}^{x}\cos\theta\sin\alpha+\sigma_{xy}^{y}\sin\theta\sin\alpha].
\end{multline}

Upon integration we find that:

\begin{equation}
E_{\theta,\alpha}=E_{x,x}\cos\theta\cos\alpha+E_{y,x}\sin(\theta+\alpha)+E_{y,y}\sin\theta\sin\alpha.\label{e_generalization}
\end{equation}

\begin{figure}[!]
\includegraphics[width=4cm]{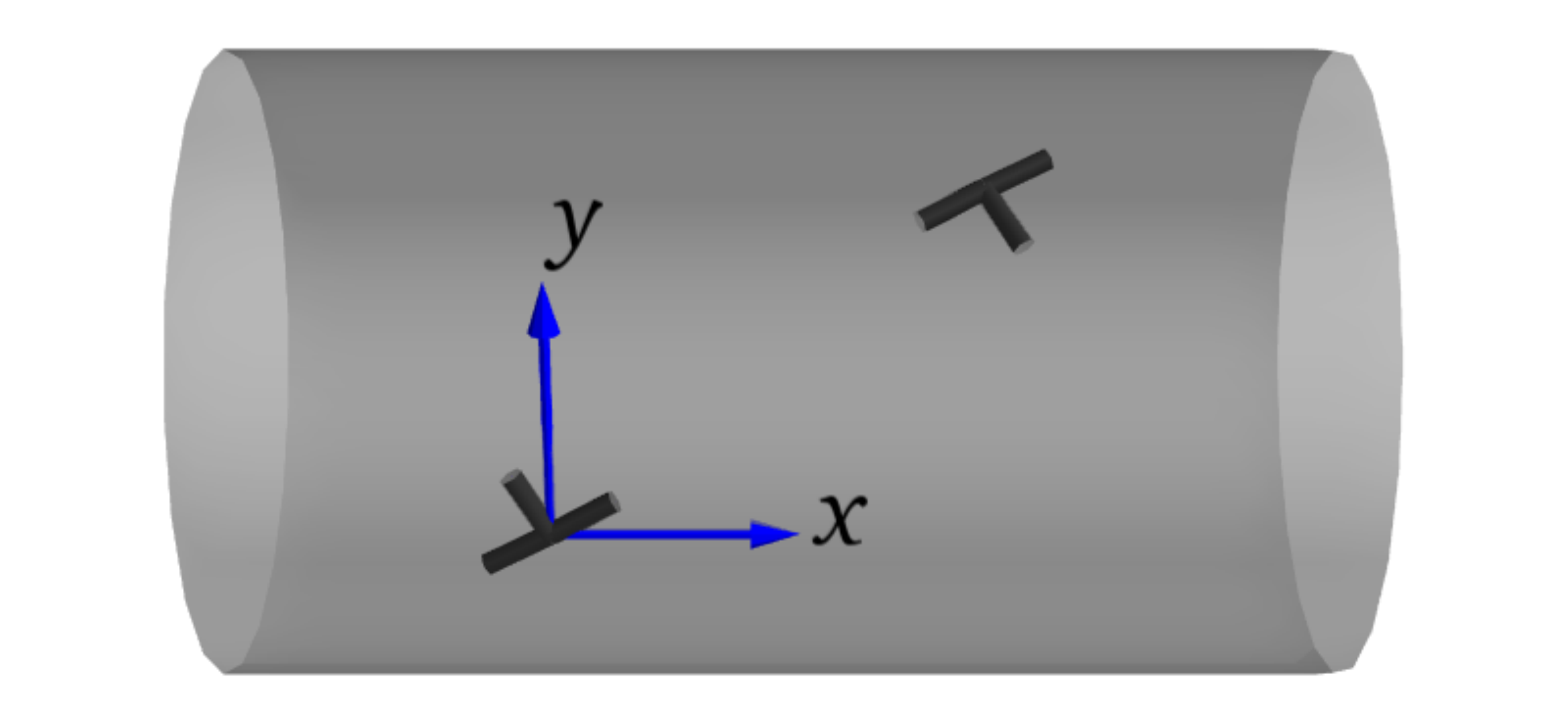}

\includegraphics[width=8cm]{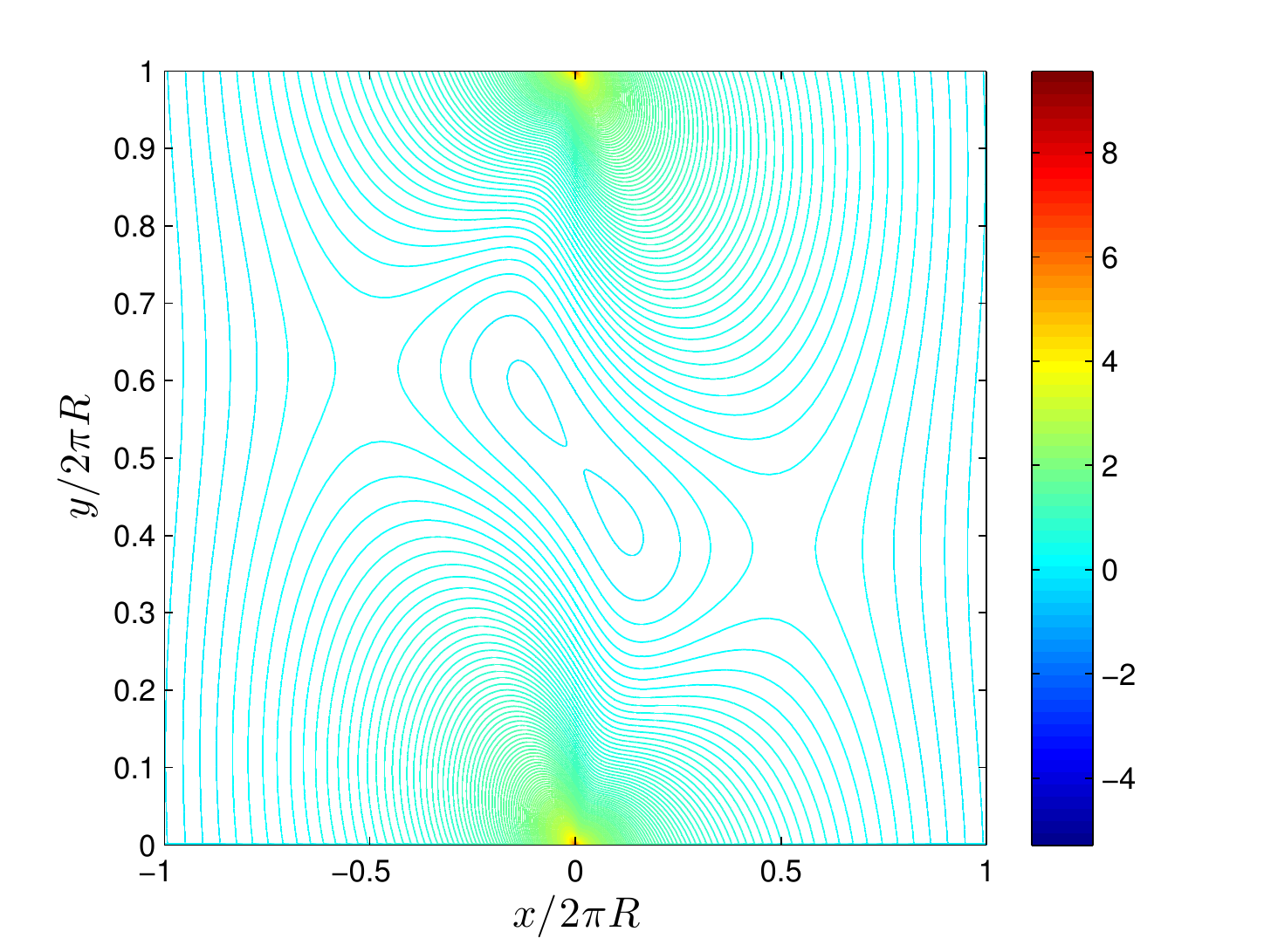} \caption{Equipotential contours for a dislocation on a cylinder with a Burgers
vector $\vec{b}=b(\frac{\sqrt{3}}{2}\hat{x}+\frac{1}{2}\hat{y})$,
forming an angle of $\pi/6$ with respect to the $\hat{x}$-axis,
interacting with another dislocation with an opposite Burgers vector.
The interaction energy is given by Eq. (\ref{e_generalization}).
Note the two skewed saddle points. Energy is measured in units of
$Ab^{2}.$ }

\label{fig_energy_triangular}
\end{figure}

For the phyllotaxis problem discussed in section \ref{phyllo}, the
energy landscape associated with dislocation pairs with antiparallel
Burgers vectors is of particular interest; these can nucleate locally
and then unbind, thus modifying the geometry of the lattice. Fig.
\ref{fig_energy_triangular} shows an example of the energy equipotential
contours for $\theta=\pi/6$ and $\alpha=7\pi/6$.

\subsection{Structure of the energy landscape}

In this section we describe in more detail the structure of the interaction
energy landscape for two dislocations interacting on the surface of
a cylinder at zero temperature.

\subsubsection{Burgers vectors are in the $\pm x$ directions}

In the case of dislocations with antiparallel Burgers vectors in the
$\pm\hat{x}$ directions, the interaction energy is given by Eq. (\ref{energy}).
From the energy contours (see Fig. \ref{contour_plot}), we can see
that the configuration where the two dislocations are located at the
same $x$ coordinate but on opposite sides of the cylinder, the anti-podal
point, is a maximum of the energy. If the two dislocations have the
same sign, there is a stable minimum riding on top of an infinite
energy due to the lack of overall charge neutrality. To see this,
one can expand the force on a dislocation offset slightly from the
origin by an amount $(x,y)$, exerted by another dislocation at the
anti-podal point. This gives, for antiparallel Burgers vectors, to
lowest non-vanishing order:

\begin{eqnarray}
F_{x}(x,y) & =Kx,\nonumber \\
F_{y}(x,y) & =Ky,\label{stability}
\end{eqnarray}

with $K=\frac{A\pi^{2}b^{2}}{W^{2}}>0,$ so the minimum is unstable.
For parallel Burgers vectors, the signs are reversed.

We conclude that two dislocations with antiparallel Burgers vectors
on opposite sides of the cylinder can minimize their energy by annihilating,
while two dislocations of the same sign will remain at the anti-podal
point to minimize the repulsive elastic interaction energy. \subsubsection{Burgers vectors are in the $\hat{x}$ and $\hat{y}$ directions}

In this case the anti-podal point is found to be a saddle point, as
can be seen near the center of Fig. \ref{energy_xy fig}. As before,
we suppress an infinite energy due to the lack of charge neutrality.

\subsubsection{Burgers vectors are in the $\pm y$ directions}

Upon repeating the same analysis for two dislocations with Burgers
vectors in the $\pm y$ directions, we find that, again, the only
extremum point is at the anti-podal point $(0,W/2)$, which Fig. \ref{energy_yy fig}
shows to be a saddle point.

\section{Numerical test on a triangular lattice}

As discussed in the Introduction, in soft matter physics a triangular
lattice of colloidal particles wrapped around a cylinder may be realizable.
If one of the principal axes of the lattice is oriented along the
circumferential direction, the minimal Burgers vector of a dislocation
can take on the following values: $\vec{b}=\pm b\hat{y},$ $\vec{b}=\pm b\left(\frac{\sqrt{3}}{2}\hat{x}\pm\frac{1}{2}\hat{y}\right)$.

\begin{figure*}
\includegraphics[width=8.6cm]{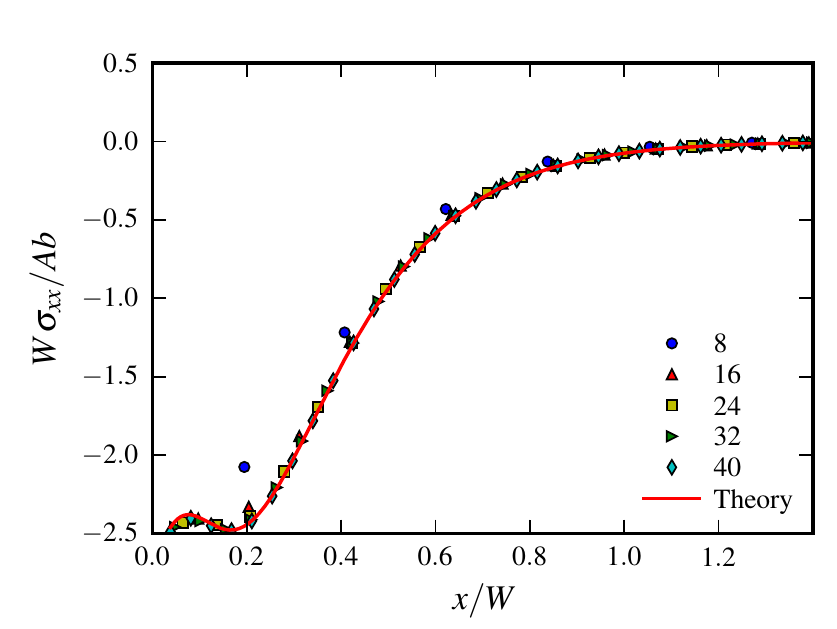}\includegraphics[width=8.6cm]{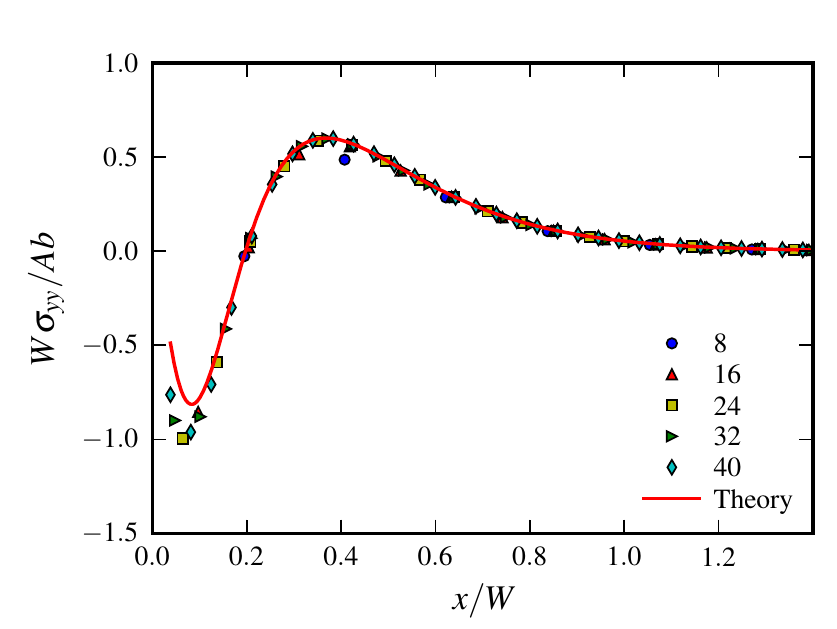}

\caption{Stress components $\sigma_{xx}$ (left) and $\sigma_{yy}$(right)
as a function of $x$ at a constant offset around the cylinder $y=W/4$,
for a dislocation with $\vec{b}=b(\frac{\sqrt{3}}{2}\hat{x}-\frac{1}{2}\hat{y)}$
at the origin. To account for a charge-neutral dislocation configuration,
another dislocation with an antiparallel Burgers vector is placed
along the negative $\hat{x}$ end of the cylinder, far enough from
the origin to create a constant $\sigma_{yy}$ stress {[}see Eq. (\ref{sigmay_yy_asympt}){]}
and a negligible contribution to the other components of the stress
tensor. The symbols are for simulations of cylinders with $W$ increasing
from $8b$ to $40b$ where $b$ is the equilibrium lattice constant
of the simulation lattice. In all cases, the length of the cylinder
is $8.66W$ and the dislocation is situated at the center of the cylinder
along the long axis to minimize edge effects. The solid line is the
theoretical prediction from Eq. (\ref{sigma_ij_theta}) with $\theta=-\pi/6$,
where for the right figure one has to add the constant $\sigma_{yy}$
stress mentioned above. Except for one or two points closest to the
dislocation, where discrete effects become important, the simulations
agree with the results of continuum elasticity even for relatively
small system sizes ($W\gtrsim16b).$ For negative values of $x$,
the convergence to the continuum limit results is slower, and there
are corrections to the stresses which scale as $\sim b/W.$ These
will be discussed in more detail in future work.}
\label{fig_simulations_sij_varyw}%

\end{figure*}

To test our continuum limit predictions numerically, consider the
stresses associated with an isolated dislocation at the origin with
Burgers vector $\vec{b}=b(\frac{\sqrt{3}}{2}\hat{x}-\frac{1}{2}\hat{y)}$
(i.e. $\theta=-\pi/6$) in a triangular lattice of masses and harmonic
springs which are relaxed to their minimum energy configuration by
a conjugate gradient method. To insure a charge-neutral configuration,
an additional dislocation with antiparallel Burgers vector was created
at the negative $\hat{x}$ end of the cylinder. The strains and stresses
at any point can be calculated from the shift in position of the surrounding
points relative to the perfect lattice. In Fig. (\ref{fig_simulations_sij_varyw}),
we display $\sigma_{xx}$ and $\sigma_{yy}$ for positive $x$ at
a constant $y=W/4$ for cylinders of various sizes. We obtain good
agreement with our continuum results even for relatively small system
sizes, which can be realized in colloidal experiments. This calculation
illustrates the applicability of our results for Burgers vectors rotated
away from the $x$-axis, which lead to long range strain fields along
the cylinder axis.

We note from Fig. (\ref{fig_simulations_sij_varyw}) that the asymptotic
value of $\sigma_{yy}$ at large positive $x$ in the simulations
is zero rather than $\sigma_{yy}^{-\pi/6}(x\to\infty)=-\pi A/W$ predicted
from Eq. (\ref{sigma_ij_theta}). This is due to the contribution
of the additional dislocation positioned at the negative $\hat{x}$
end of the cylinder: from Eqs. (\ref{sigmax_yy_asympt}), (\ref{sigmax_xx_asympt}),(\ref{sigmax_xy_asympt})
we see that while it has an exponentially small effect on the other
components of the stress tensor, it creates a \textit{constant} circumferential
stress $\sigma_{yy}$, thus shifting the values for the circumferential
stress by a constant. At $x\gg W$, the stress created by the dislocation
at the origin is also approximately constant, and since the two dislocations
are antiparallel their contributions are equal but of opposite sign
-- which explains why $\sigma_{yy}=\sigma_{yy}^{-\pi/6}+\sigma_{yy}^{5\pi/6}\rightarrow0$
at large positive $x.$

There is some evidence for helical motion of tracer particles on the
cell walls of elongating bacteria such as \textit{Escherichia coli}
\cite{wang_helical}, suggesting that the lattice shown in Fig. \ref{cylinder}
may be slightly skewed. Hence, it is of some interest to find the
asymptotic form as $x\rightarrow\infty$ of the stress for a dislocation
with a Burgers vector in a direction forming a small angle $\theta$
with the $\hat{x}$ axis. We use Eqs. (\ref{sigmay_yy_asympt}) and
(\ref{sigma_ij_theta}) to find:
\begin{multline}
\sigma_{yy}\approx\theta\cdot2\pi Ab/W\cdot Sg(x)\\
+4\pi^{2}Abe^{-2\pi|x|/W}|x|/W^{2}{\rm [{sin}(2\pi y/W)-\theta{\rm {cos}(2\pi y/W)}]}.\label{sigmax_yy_small_angle}
\end{multline}

Thus, for slightly tilted Burgers vectors there is a length scale
$l^{*}$ along $\hat{x}$ at which the constant contribution to the
stress dominates over the exponentially decaying one, which is readily
found by equating the two terms:

\begin{equation}
[2\pi e^{-2\pi l^{*}/W}l^{*}/W]{sin}(2\pi y/W)=\theta.
\end{equation}

Provided $y$ is not too small, with $W=2\pi R$, we find that to
leading order:

\begin{equation}
l^{*}\sim{\rm R{\rm log}(}\theta).
\end{equation}

\section{Analogy with grain boundaries}

\label{analogy_grain}

We now describe a correspondence between dislocations on a cylinder
and grain boundaries, which allows checks of some of our calculations
\cite{irvine_comment}.

Suppose that the lattice orientation on the cylinder allows two dislocations
of opposite Burgers vectors, $\pm b\hat{x}$, initially very close
together. Let us now glide one of the dislocations away from the other,
until they reach a certain separation $\Delta$ along the $\hat{x},$
which coincides with the axis of cylindrical symmetry. Every time
the dislocation glides by one lattice spacing, we connect an initially
circumferential row with an adjacent one to make a spiral. Thus, at
any fixed separation, the finite part of the cylinder between the
two dislocations is converted to a \textit{helical} structure, which
can be thought of as a local rotation of the crystalline lattice (see
Fig. \ref{magnetic}). Since a shift of $b$ is associated with each
rotation, the pitch of the helix is $\theta=b/W$. Therefore, we can
view the circumferential row passing through each of the dislocations
as a \textit{grain boundary}, and the whole structure as a polycrystalline
material with three regions, the middle one tilted by $\theta$. Glide
separation and the equivalent grain boundary configuration are also
illustrated in Fig. \ref{grain}, as well as in the Supplementary
Video S1, showing numerical results for a system of masses and springs,
where the minimal energy configuration of the system at each instance
in time is obtained by a conjugate gradient minimization procedure.

\begin{figure*}
\includegraphics[width=17cm]{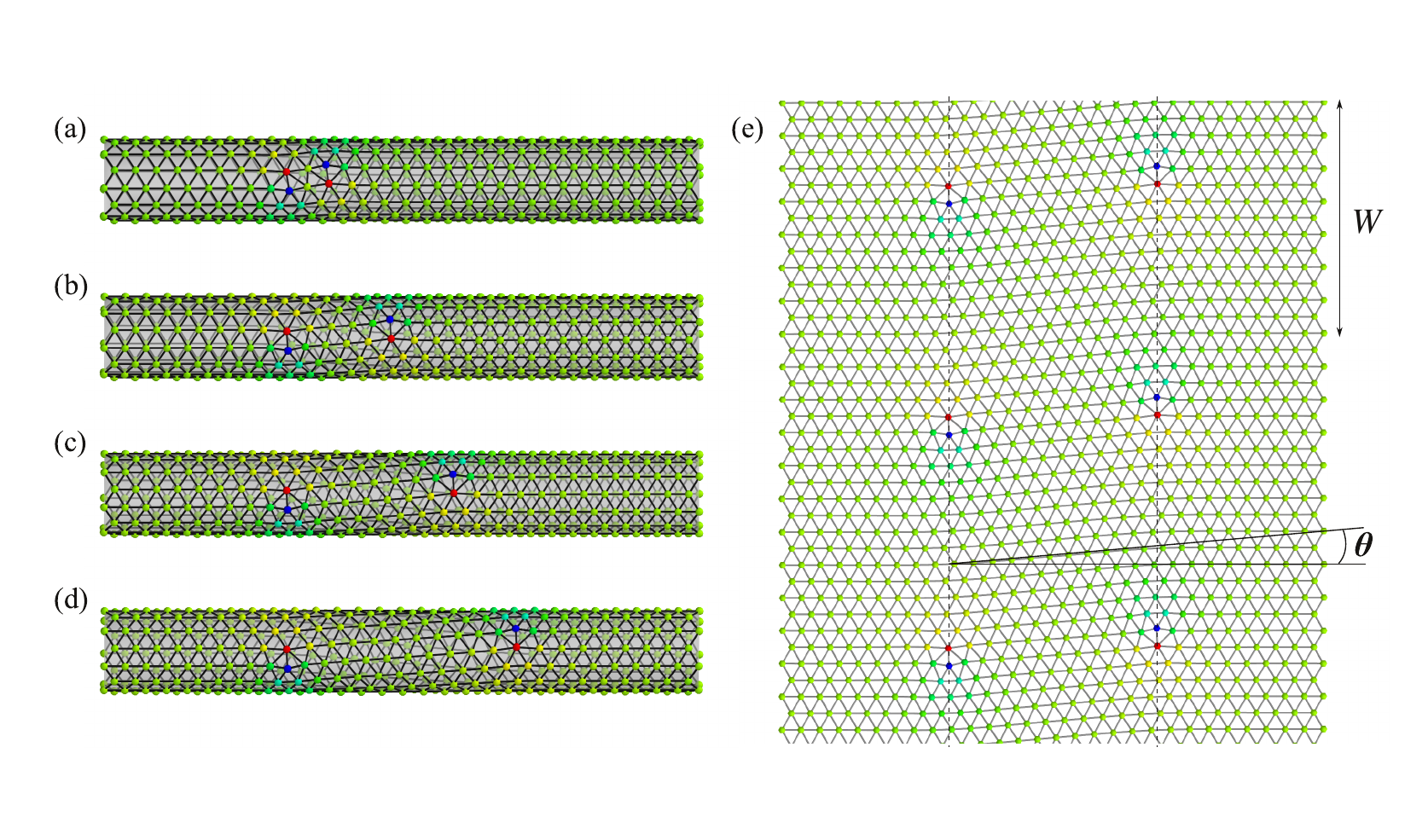}\caption{Dislocations on a cylinder are analogous to grain boundaries. The
images show configurations obtained numerically for a hexagonal lattice
of masses and harmonic springs whose energy is minimized as the dislocation
pair is separated by a glide. On this triangular lattice, a dislocation
corresponds to a point with five neighbors adjacent to a point with
seven neighbors. Starting with two nearby dislocations, (a), we glide
one of the dislocations to the right in steps of the lattice spacing
{[}Supplementary Video S1; figures (b--d) show intermediate snapshots
during the glide{]}. The finite section of the cylinder between the
two dislocations displays a helical structure, similar to the model
with magnetic beads shown in Fig. \ref{magnetic}. The analogy with
a \emph{flat }polycrystalline material is illustrated by unrolling
the cylindrical crystal in (d) and duplicating it several times in
the $y$ (circumferential) direction to replicate the periodic boundary
conditions (e). The resulting infinite columns of dislocations (dotted
lines) define two grain boundaries separating the middle region, rotated
by an angle $\theta=b/W$, from sections with the original crystal
orientation on either side.}

\label{grain}
\end{figure*}

By this analogy, the interaction energy $E(x,y)$ of two dislocations
on the cylinder separated by a displacement $(x,y)$, corresponds
to the energy (per section of length $W$) of two infinite, straight,
grain boundaries a distance $x$ apart and offset by an amount $y$
along their lengths, where the distance between adjacent dislocations
in each grain boundary is $W$. As is well known in the theory of
small angle grain boundaries \cite{hirth,bruinsma}, the associated
grain boundary angle is $\theta=b/W.$ For example, the interaction
energy of two parallel grain boundaries with Burgers vectors in the
$\pm\hat{x}$ directions and separated by a distance $x$, mathematically
equivalent to the situation in Fig. \ref{grain}(e), is given by Eq.
(\ref{eq:limit_bruinsma}) ; This is in agreement with Eq. (2.1b)
in Ref. \cite{bruinsma}.

\section{energy landscape in the presence of a constant azimuthal force}

Consider two dislocations with antiparallel Burgers vectors $\pm b\hat{x}$
on a cylinder, a small distance apart. For simplicity, we pin one
dislocation at the origin. Let us assume that there is an additional
constant force acting on it, due, say, to the Peach-Koehler force
created by an external stress $\sigma_{xx}$ {[}see Eq. (\ref{force_y}){]}.

The energy landscape now resembles a tilted washboard potential with
period $W$ and is given by:

\begin{multline}
E_{\hat{x},-\hat{x}}^{F}(x,y)=\frac{Ab^{2}}{2}{\rm {log}}[{\rm \frac{W}{\pi b}sinh}(\pi(x-iy)/W)]+{\rm }\\
+\frac{Ab^{2}}{2}i\pi(x/W){\rm {csc}(\pi y/W){\rm {sinh}(\pi x/W){csch}(\pi(x-iy)/W)}}\\
+C.C.-Gy,\label{energy-1}
\end{multline}
where the constant azimuthal force $G\propto\sigma_{xx}.$ For climb
dynamics, there can also be a contribution due to the chemical potential
associated with adding new material \cite{amir_nelson_pnas}. The
dynamics of the dislocation pair in this potential will depend on
the glide and climb mobilities to be introduced below. First, however,
we discuss how the energy landscape changes due to the field $G$.
Clearly, the minimum at the origin remains if the field $G$ is not
too large. Now, however, there are important new \textit{saddle points}
in the energy landscape. Let us first locate these saddles in the
case of flat space. Here, the potential is given by Eq. (\ref{energy_flat}),
with an additional term $-Gy$. A necessary condition for the existence
of a saddle is the vanishing of $\frac{\partial E}{\partial x}$ and
$\frac{\partial E}{\partial y}$. Upon differentiation of $E_{\hat{x},-\hat{x}}^{F}(x,y)$
we obtain:

\begin{equation}
x_{saddle}=y_{saddle}=\frac{Ab^{2}}{G}.\label{saddle_flat}
\end{equation}
 This extremum must be a saddle point since (a) for a given $y$ coordinate
there is a minimum energy at $x=y$ {[}see discussion preceding Eq.
(\ref{x-vanish}){]}, and (b) this cannot be a minimum in the $y$
direction since $y\rightarrow0$ and $y\rightarrow\infty$ give us
a negative, diverging energy.

Let us now consider the problem on a cylinder. The only scales that
may enter the problem are $Ab^{2},W$ and $G$; $b$ cannot enter
explicitly since we are considering the continuum limit. Only one
dimensionless parameter can be formed out of these, namely: $D\equiv GW/Ab^{2}$.
Therefore the $y$ position of the saddle (which as we show below
exists for any value of $G$), can be written as:

\begin{equation}
y_{saddle}=\frac{Ab^{2}}{G}f(D),\label{eq:scaling}
\end{equation}
where $f(D)$ is a function of the dimensionless parameter $D.$ The
case of flat space, Eq. (\ref{saddle_flat}), corresponds to $D\rightarrow\infty$,
showing that $f(D)$ must asymptotically approach the value one. We
now proceed to discuss the general form of $f(D)$. The condition
$\frac{\partial E}{\partial x}$ does not depend on the field in the
$y$ direction, and thus, Eq. (\ref{x-vanish}) still holds. Therefore,
we search for the saddles whose coordinates have the form $(x^{*}(y),y)$,
where:
\begin{equation}
x^{*}(y)=\pm\frac{W}{\pi}{\rm {arctanh}({\rm {tan}(\pi y/W)),}}\label{xstar}
\end{equation}
 with $0<y<W/4$.

\begin{figure}
\includegraphics[width=8cm]{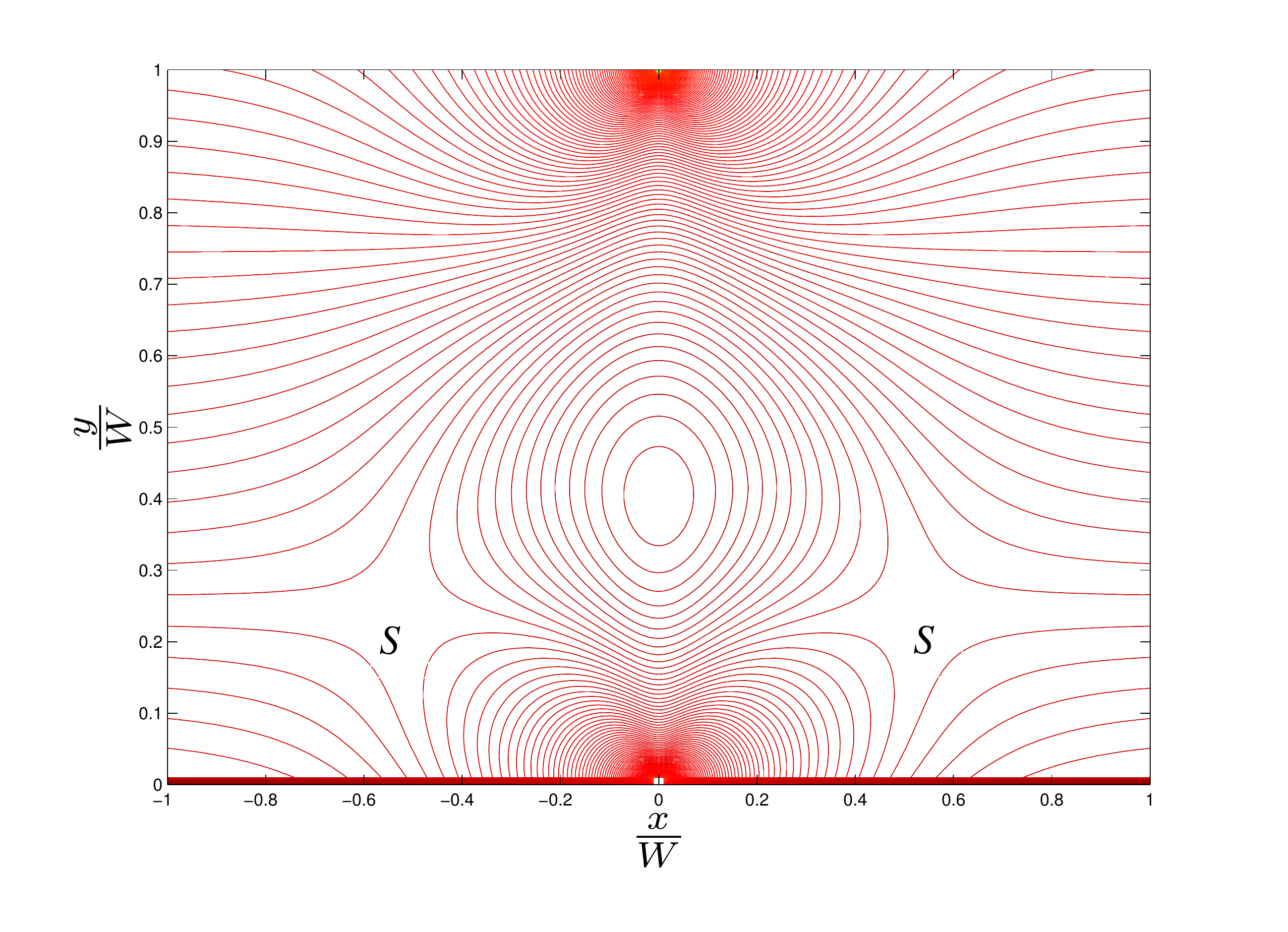}

\caption{Equipotential contours for a dislocation pair with$\frac{GW}{Ab^{2}}=1,$
with $W=2\pi R$. Note that the force $G$ caused by a macroscopic
stress $\sigma_{xx}$ has pulled the maximum away from $y=W/2$. Note
also the two saddle points (denoted by the letter \textit{S}) near
$y\approx W/4.$}

\label{saddle_1}
\end{figure}

\begin{figure}
\includegraphics[width=8cm]{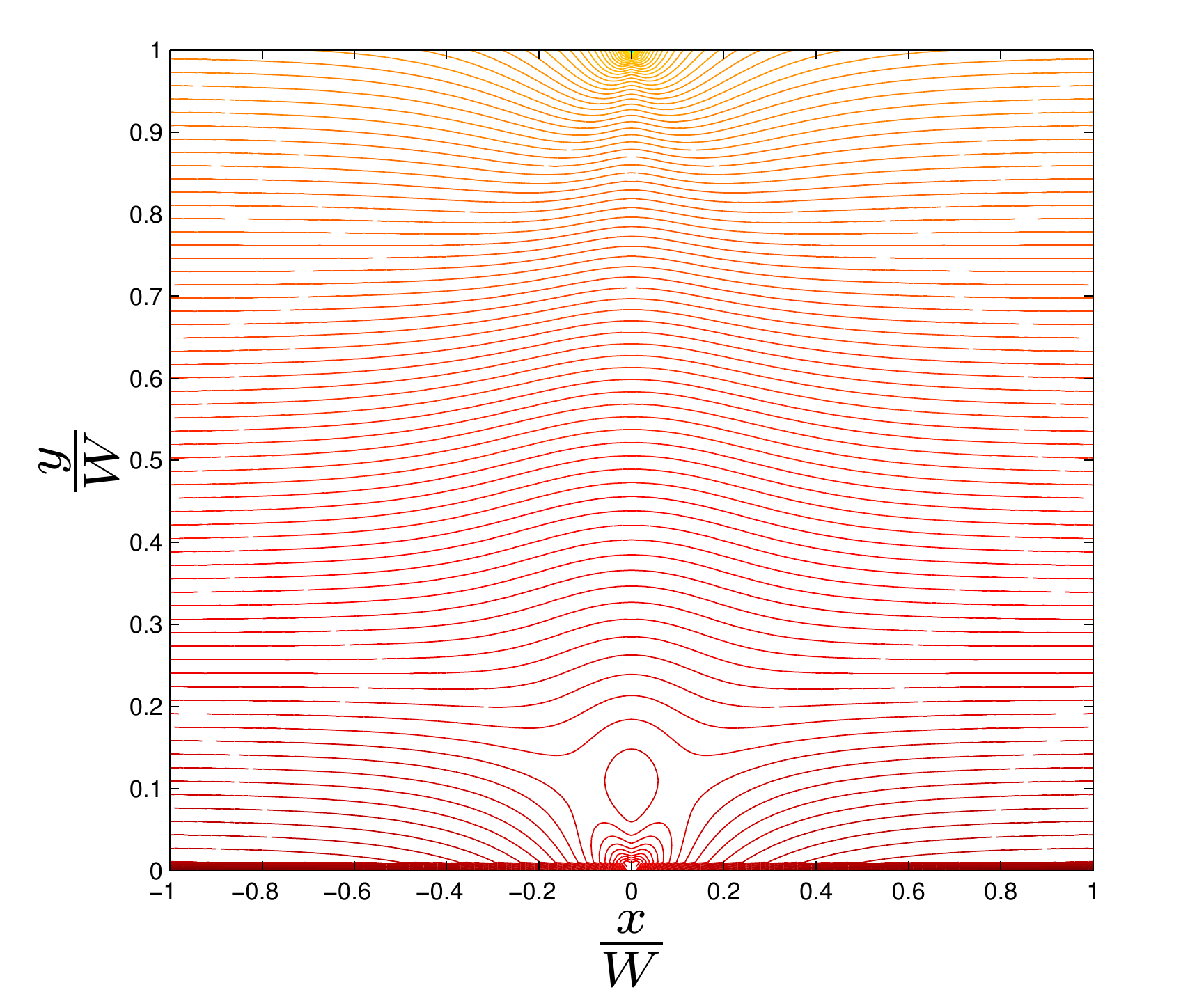}

\caption{Equipotential contours for a dislocation pair with $\frac{GW}{Ab^{2}}=10$.
Note the maximum and two saddle points near the line $y=W/10.$}

\label{saddle_2}
\end{figure}

\begin{figure}
\includegraphics[width=8cm]{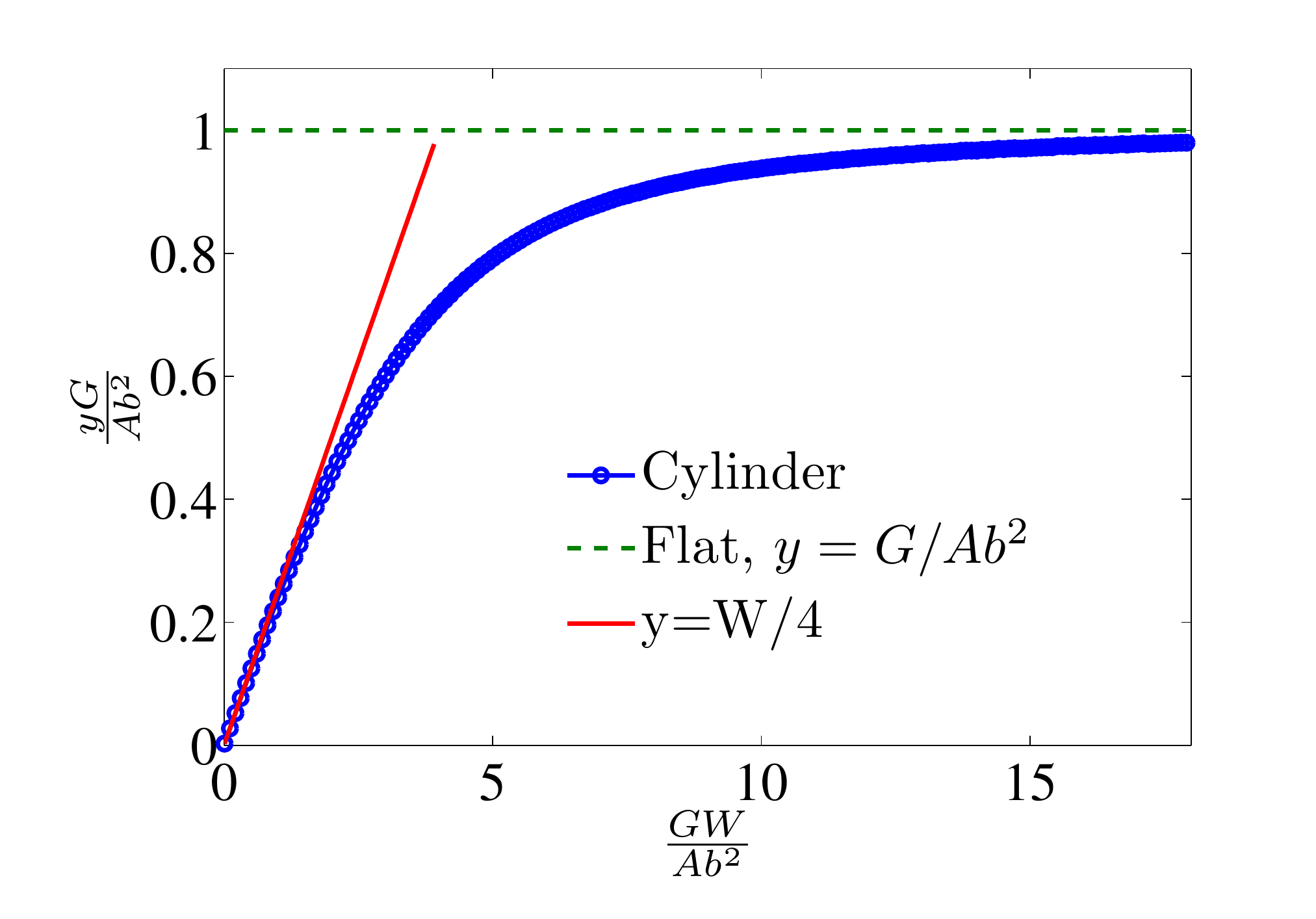}

\caption{The rescaled $y$ coordinate of the saddle point $\frac{yG}{Ab^{2}}$
is shown as a function of the dimensionless parameter $\frac{GW}{Ab^{2}}$.
The horizontal line is the result in flat space, where $y=\frac{Ab^{2}}{G}.$}

\label{saddle_y}
\end{figure}

Within this parametrization, we still have to satisfy $\frac{\partial E}{\partial y}=0$.
This condition implies that $F_{y}=G-b\sigma_{xx}^{x}=0$, with $\sigma_{xx}^{x}$
given by Eq. (\ref{sigma_xx_cylinder-x}). For small values of $y$,
a vanishing derivative in the $x$ direction would yield $x^{*}(y)=y$,
as can also be seen from Eq. (\ref{xstar}). For such a point, we
know that the two dislocations have a large force in the $y$ direction
pulling them together, which can be seen directly from Eq. (\ref{sigma_xx})
(since in this regime we are not sensitive to the cylindrical geometry).
On the other hand, for $y$ very close to $W/4$, Eq. (\ref{xstar})
yields an $x$ coordinate which diverges. Hence, the stress in Eq.
(\ref{sigma_xx_cylinder-x}) must vanish, since the force between
two dislocations falls off as a power-law. This argument implies that
the $y$ component of the force at such a point will be approximately
$G$, and in particular, it will be \emph{positive}. Therefore, there
is an intermediate value of $y$ for which both $F_{x}$ and $F_{y}$
vanish. We conclude that a saddle exists for any value of $G$! (In
fact, two saddles, since we have a reflection symmetry around the
$y$ axis.) Equipotential contours where the saddles can be seen are
shown in Figs. (\ref{saddle_1}) and (\ref{saddle_2}), for values
$\frac{GW}{Ab^{2}}=1$ and a much stronger force given by $\frac{GW}{Ab^{2}}=10$,
respectively. For large values of $G$, the saddles coincide with
those obtained for flat space. However, for small $G$ the saddles
will be close to the line $y=W/4$ and their $x$ values will \textit{diverge}.
Fig. (\ref{saddle_y}) shows this scaling, expressed in Eq. (\ref{eq:scaling}),
and the asymptotic limits: for $GW/Ab^{2}\rightarrow\infty,$ the
rescaled $y$ coordinate of the saddle $\frac{yG}{Ab^{2}}$ approaches
unity, corresponding to the flat space result. For the opposite case,
$\frac{yG}{Ab^{2}}\approx\frac{WG}{4Ab^{2}}$, corresponding to the
straight line $f(D)=D/4.$ To determine the form of the divergence
of the $x$ position of the saddle for small $G$, note from Eq. (\ref{sigmax_xx_asympt})
that at large values of $x$ the force in the $y$ direction decays
exponentially for any value of $y$, as:

\begin{equation}
F_{y}\approx[4\pi^{2}Ab^{2}e^{-2\pi x/W}x/W^{2}{\rm ]{sin}(2\pi y/W).}\label{FyAsymptotic}
\end{equation}
Upon equating this force to $G$, we find for small $G$ a saddle
at $x*\sim-\frac{W}{2\pi}{\rm log\left(\frac{GW}{Ab^{2}4\pi{\rm {\rm {log}}\left(GW/Ab^{2}\right)}}\right)}$,
a location that diverges approximately logarithmically as $G\rightarrow0$.

~
~

\begin{figure}
\includegraphics[width=8cm]{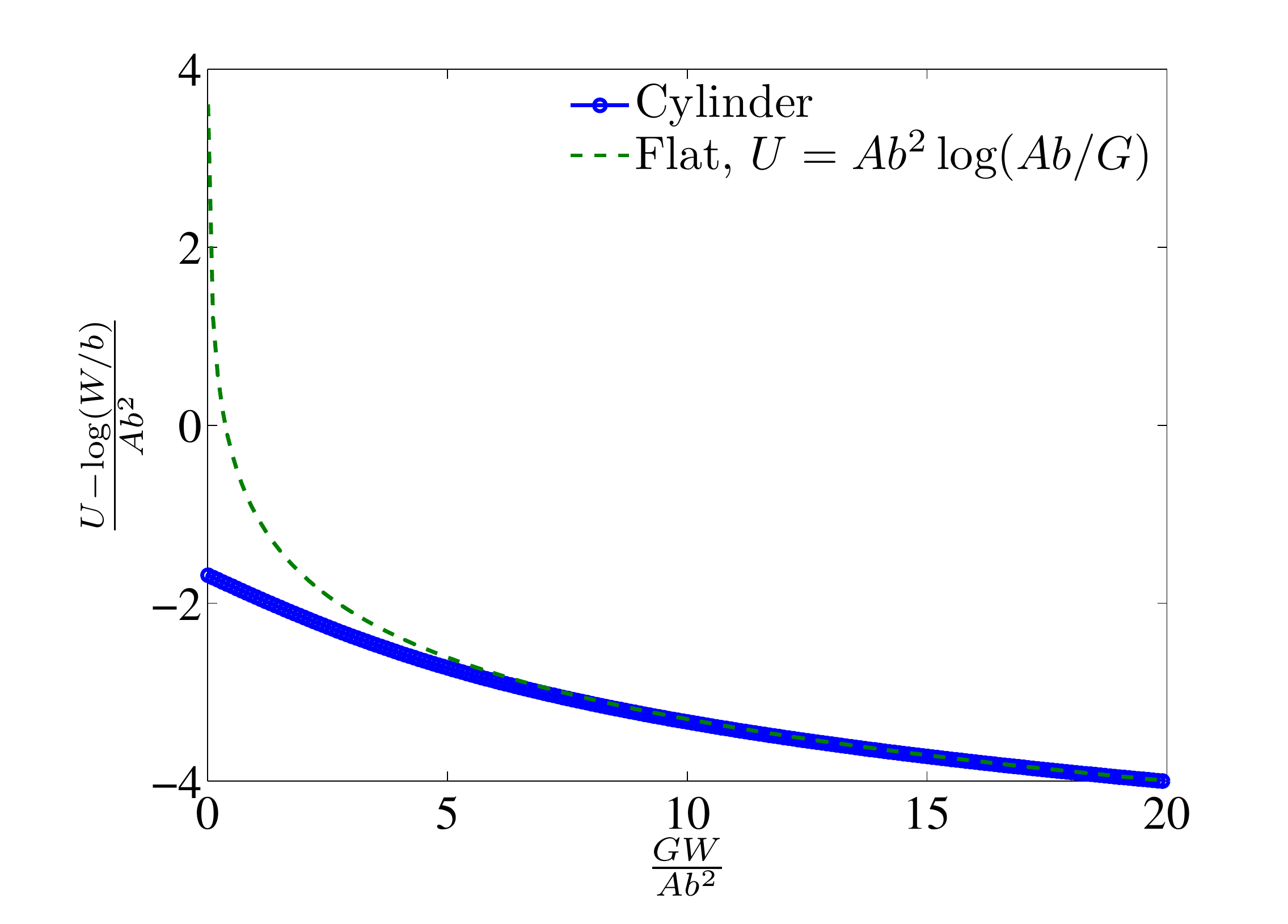}

\caption{The saddle point excitation energy $U$, given by Eq. (\ref{energy_xx}),
is rescaled and plotted as a function of $\frac{GW}{Ab^{2}}$. In
the flat space limit the scaling function is given by $\eta(D)=-\log(D),$
corresponding to $U=Ab^{2}\log(Ab/G)$. For the opposite case, when
$G$ is small and the cylindrical geometry is important, the energy
$U$ saturates at a constant value even as $G\rightarrow0$, $\frac{U}{Ab^{2}}\approx const+\log(\frac{W}{b})$. }

\label{saddle_energy}
\end{figure}

An important feature of the saddle when considering the thermal excitation
of dislocation pairs is its energy relative to, say, the interaction
energy of the two dislocations when they are one lattice spacing apart.
This energy difference is plotted in Fig. \ref{saddle_energy}. For
flat space, it is easy to see that the energy barrier diverges logarithmically
as $G\rightarrow0$, and is given by:

\begin{equation}
U(G)\approx Ab^{2}\log(Ab/G),\label{E_flat}
\end{equation}
 where the energy is measured relative to that of a dislocation pair
separated by a lattice constant $b$.

Eq. (\ref{E_flat}) is approximately correct on the cylinder as well,
provided the $y$ coordinate of the saddle is much less than $W$.
For small values of $G$, more careful analysis is required. Let us
parametrize the saddle point by $(x^{*}(y),y).$ From Eq. (\ref{sigma_xy_cylinder-x})
we infer that at large values of $x$ the force in the $x$ direction
decays exponentially for any value of $y$, as:

\begin{equation}
F_{x}\approx[4\pi^{2}Ab^{2}e^{-2\pi x/W}x/W^{2}]{\rm {cos}(2\pi y/W).}\label{FxAsymptotic}
\end{equation}
Hence, the potential at infinity does not diverge, leading to a finite
energy barrier for dislocation unbinding at any value of $G$, as
shown in Fig. (\ref{saddle_energy}). We can adapt the scaling analysis
for the $y$ coordinate of the saddle also to the energy. Consider
the interaction energy $E(x,y)$ in the absence of $G$$.$ Although
the lattice spacing $b$ enters the expression for the interaction
energy explicity, it can be easily eliminated: from Eq. (\ref{energy_xx})
we see that

\begin{equation}
E(x,y)=Ab^{2}\log(W/b)+Ab^{2}h[x/W,y/W],\label{eq:E_scaling}
\end{equation}

which can also be deduced from dimensional considerations combined
with the flat space limit. The energy $U(G)$ is thus given by:

\begin{equation}
U(G)=E(x_{saddle},y_{saddle})-E(|\vec{r}|=b)-Gx_{saddle}.\label{eq:U}
\end{equation}
Previously we found that the $y$ coordinate of the saddle point scales
as $y_{saddle}=\frac{Ab^{2}}{G}f(D),$ see Eq. (\ref{eq:scaling}).
The same argument can be repeated for the $x$ coordinate of the saddle,
giving $x_{saddle}=\frac{Ab^{2}}{G}g(D),$ where the function $g$
also depends on the dimensionless parameter $D.$ Plugging these results
into Eq. (\ref{eq:U}) and using Eq. (\ref{eq:E_scaling}) we find
that:

\begin{equation}
U(G)=Ab^{2}\eta(D)-Ab^{2}\log(b/W),
\end{equation}
with $\eta(D)=h[\frac{g(D)}{D},\frac{f(D)}{D}]-g(D)$ a function of
the dimensionless parameter $D$. Rearranging the above equation,
we find a dimensionless scaling form:

\begin{equation}
\frac{U(G)-Ab^{2}\log(W/b)}{Ab^{2}}=\eta(D),
\end{equation}
For flat space, we have $U(g)=Ab^{2}\log(Ab/G)$, showing that for
large values $\eta(D)\approx-\log(D)$. The above form of scaling
is illustrated in Fig. \ref{saddle_energy}.

\section{Thermal nucleation of dislocation pairs: a two-dimensional escape over a barrier problem}

With colloidal particle arrays or bacterial cell walls at relatively
high temperatures in mind, we shall now calculate the thermal nucleation
rate of dislocations: in the presence of a uniform external field
denoted $G$, two dislocations of opposite Burgers vectors can overcome
the energy barrier due to their attractive interaction and unbind
due to thermal fluctuations. We shall first treat this process in
flat space, and then compare to the results in cylindrical geometry,
which differ significantly when $G$ is small. Our analysis will be
done for the case of dislocations with Burgers vectors $\pm b\hat{x},$
where the external force driving the nucleation is in the $\hat{y}$
direction, perpendicular to the Burgers vectors. For simplicity, we
assume isotropic elastic constants. The same strategy can be adapted
to other scenarios. We then discuss the nucleation of dislocations
along the cylinder axis due to a \emph{twisting} stress $\sigma_{xy}$.

\subsection {Nucleation of dislocations in flat space}

We consider first the thermal escape-over-a-barrier problem for two
dislocations in flat space (similar results should hold for a cylinder
with a large enough field $G$ such that the saddle distance from
the origin is much smaller than $W$). It is convenient to assume
that one of the dislocations is mobile, while the other is pinned
at the origin. Alternatively, we can consider the motion relative
to the fixed center-of-mass of the dislocation pair. For a similar
analysis of thermal activation of vortex pairs in superfluid helium
films, see Ref. \cite{vortex_nucleation}. Throughout this section,
we choose units such that the Boltzmann constant $k_{B}=1$.

We shall assume over-damped dynamics for the mobile dislocation, described
by an anisotropic diffusion tensor ${\bf D}$ \cite{zippelius}. We
analyze the scenario where the dislocations can \textit{glide} and
\textit{climb}. This choice is motivated by the bacterial elongation
problem discussed in section \ref{bacteria}, where it is the climb
of dislocations that drives the growth process. The motion of dislocations
for colloidal arrays on a cylinder could also have a climb component,
provided the particles can jump on and off the cylinder, thus providing
an external source of vacancies and interstitials. Notice that allowing
climb mobility is very different from the situation in Ref. \cite{bruinsma},
where dislocations can only glide and one has a one-dimensional escape-over-a-barrier
problem, rather than the \textit{two-dimensional} problem that we
study here.

The probability $n(\vec{r},t)$ of finding the dislocations with separation
$\vec{r}$ is given by the continuity equation:

\begin{equation}
\frac{dn}{dt}+div\vec{j}=0,
\end{equation}
 with a probability current $\vec{j}(\vec{r},t)$:

\begin{equation}
\vec{j}=\boldsymbol{\mu}n\vec{F}(\vec{r})-{\bf {D}}\vec{\nabla}{n}.\label{currents}
\end{equation}

The first term is given by $\vec{F}(\vec{r})=-\vec{\nabla}U(\vec{r})$,
where the potential energy $U(\vec{r})$ is the sum of the interaction
of the two dislocations as given by Eq. (\ref{energy_xx}) and the
potential $-Gy$ associated with the external driving force. Sufficiently
close to equilibrium, the Einstein relation relates the diffusion
tensor to the mobility tensor: ${\bf D}=T{\bf \boldsymbol{\mu}}$,
allowing us to recast the Fokker-Planck equation as (setting $k_{B}=1$):

\begin{equation}
\frac{dn}{dt}=-\vec{\nabla}[-{\bf D}e^{-U/T}\vec{\nabla}(ne^{U/T})].
\end{equation}
 We now exploit Langer's formalism \cite{langer} to determine the
structure of the probability currents that describe the nucleation
of dislocation pairs.

For small enough escape rates (\textit{i.e.}, a low enough temperature),
the process will be dominated by currents flowing near the saddles
points of the potential energy landscape. In the case of a dislocation
pair in flat space there are two symmetric saddles; the total escape
rate will be double the probability current flowing through one of
them.

The escape rate through the saddle includes the Hessian (matrix of
second derivatives) around it. We denote the Hessian in the vicinity
of the saddle by ${\bf H_{saddle}}$, and the Hessian in the vicinity
of the minimum by ${\bf H_{min}}$(for dislocation pairs, the latter
requires a short distance cutoff of order the lattice spacing $b$).
The probability density around the minimum follows a Boltzmann distribution,
since thermalization in that region is a fast process compared to
the time scales associated with crossing the high barriers. It is
useful to define the transition matrix ${\bf A}$ as:

\begin{equation}
{\bf A}={\bf D}{\bf H_{saddle}},
\end{equation}
 and denote its two eigenvalues by $\lambda_{1}>0$ and $\lambda_{2}<0$.

The escape rate $\Gamma$ is then given by \cite{langevin} :

\begin{equation}
\Gamma=\frac{\lambda_{1}}{2\pi}\sqrt{\frac{{det}[{\bf H_{min}}]}{|{det}[{\bf H_{saddle}}]|}}e^{-U_{saddle}/T}.\label{langer}
\end{equation}
We shall assume that ${\bf D}$ is anisotropic but is diagonal in
a basis where one of the eigenvectors coincides with the direction
of the Burgers vector, say, $\hat{x}$. In this basis:

\begin{equation}
{\bf {\bf D}}=\left(\begin{array}{cc}
D_{g} & 0\\
0 & D_{c}
\end{array}\right),\label{eq:diffusion}
\end{equation}
 where $D_{c}$ and $D_{g}$ are the glide and climb diffusion coefficients.
As mentioned before, for bacterial cell wall growth dislocation motion
is predominantly via climb, \textit{i.e.}, $D_{c}\gg D_{g}.$ For
the case of colloids on a cylinder, we would typically have glide
dynamics, \textit{i.e.}, $D_{g}\gg D_{c}.$

Upon finding the positions of the two symmetric saddle points and
calculating their Hessians, we obtain:

\begin{equation}
{\bf H_{saddle}}=\frac{G^{2}}{2Ab^{2}}\left(\begin{array}{cc}
1 & -1\\
-1 & -1
\end{array}\right)
\end{equation}
 It follows that $\sqrt{|{det}[{\bf H_{saddle}}]|}=\frac{G^{2}}{\sqrt{2}Ab^{2}}$,
and $\lambda_{1}=\frac{G^{2}}{Ab^{2}}[\frac{D_{g}-D_{c}}{2}+\frac{1}{2}\sqrt{(D_{c}-D_{g})^{2}+8D_{c}D_{g}}]$.

Eq. (\ref{langer}) then leads to an escape rate given by:

\begin{equation}
\Gamma\propto\left[\frac{D_{g}-D_{c}}{2}+\frac{1}{2}\sqrt{(D_{c}-D_{g})^{2}+8D_{c}D_{g}}\right]e^{-U_{saddle}/T},\label{rate_flat}
\end{equation}
 with $U_{saddle}=Ab{\rm ^{2}{log}(\frac{Ab^{2}\sqrt{2}}{G})+C}$,
according to Eq. (\ref{energy_xx}), with $C$ a constant depending
on the non-universal details associated with ${det}[{\bf H_{min}}].$

The proportionality constant for the rate in Eq. (\ref{rate_flat})
will depend on the details of the dislocation nucleation at the origin,
which should be independent of $G$. We conclude that the escape rate
goes to zero as a power-law as $G\rightarrow0$:
\begin{equation}
\Gamma\propto\left(\frac{G}{Ab}\right)^{\frac{Ab}{T}}.\label{eq: flat}
\end{equation}

\subsection {Nucleation of dislocations on a cylinder}

We now solve the escape problem on a cylinder. The result for flat
space should hold for large enough $G$ so that $\frac{Ab^{2}}{G}\ll W$.
When this inequality is reversed, however, the saddles will occur
(as discussed above) at a point with large $x$ and $y\approx W/4$,
with $U_{saddle}$ approaching a finite, maximal value $U_{barrier}$,
independent of $G$ (see Fig. \ref{saddle_energy}). Consider the
form of ${\bf H_{saddle}}$ in this regime. From Eqs. (\ref{sigmax_xy_asympt})
and (\ref{sigmax_xx_asympt}) we know the asymptotic form of $\frac{\partial E}{\partial x}$
and $\frac{\partial E}{\partial y}$ for large $x$. %tested in test_asymptotic.m
With this information we can calculate the Hessian,

\begin{equation}
{\bf H_{saddle}}=\frac{G^{2}}{Ab^{2}}\left(\begin{array}{cc}
-\frac{\partial F_{x}}{\partial x} & -\frac{\partial F_{x}}{\partial y}\\
-\frac{\partial F_{x}}{\partial y} & -\frac{\partial F_{y}}{\partial y}
\end{array}\right).
\end{equation}
A straightforward calculation yields:

\begin{equation}
{\bf H_{saddle}}=[8\pi^{3}Ab^{2}2\pi x/W^{3}e^{-2\pi x/W}]{\bf M}(y),
\end{equation}

where the matrix ${\bf M}$ is given by the $y$ coordinate of the
saddle:

\begin{equation}
{\bf M}=\left(\begin{array}{cc}
{\rm {cos}(2\pi y/W)} & {\rm {sin}(2\pi y/W)}\\
{\rm {sin}(2\pi y/W)} & -{\rm {cos}(2\pi y/W)}
\end{array}\right).
\end{equation}

\begin{figure}
\includegraphics[width=8cm]{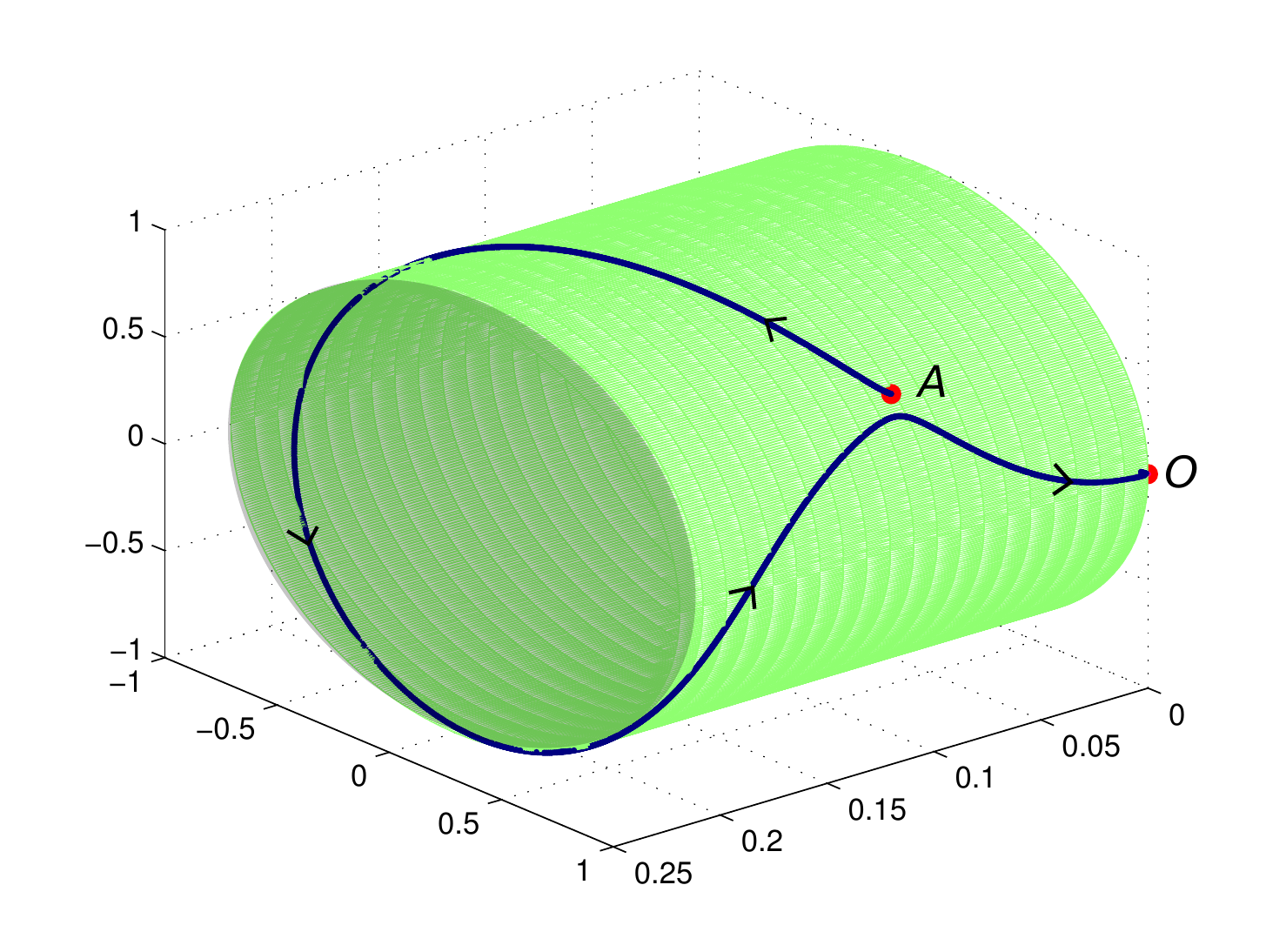}

\caption{A mobile dislocation, initially at point \textit{A}, interacting with
a pinned dislocation with antiparallel Burgers vector at the origin
$O$. The first dislocation is placed just beyond the saddle point,
at $(x,y)=(\frac{W}{10},\frac{W}{10})$, for $\frac{GW}{Ab^{2}}=10$
(for a flat space, this would be the exact position of the saddle,
however on a cylinder this point lies beyond it). The temperature
is close to zero, although we assume motion over the Peierls potential
\cite{hirth} is possible. Here, $W=2\pi R=1$, and the isotropic
mobility tensor components are $\mu_{c}=\mu_{g}=1$. After completing
one loop, the dislocation is ``recaptured'' and annihilates with
the dislocation at the origin.}

\label{escape_prob}

\end{figure}

Since $y\approx W/4$, we have:
\begin{equation}
{\bf M}\approx\left(\begin{array}{cc}
0 & 1\\
1 & 0
\end{array}\right),
\end{equation}
 Using Eq. (\ref{langer}) we therefore obtain that:

\begin{equation}
\Gamma\propto\sqrt{D_{c}D_{g}}e^{-E_{barrier}(G)/T}.\label{rate_cylinder}
\end{equation}
 Thus, as the field $G$ becomes weaker, and the cylindrical geometry
becomes more important, the escape rate decreases but saturates at
a finite value! Note, however, that the escape problem on a cylinder
does not fit precisely to the conditions of the Kramers problem: one
of the usual assumptions is that once the particle escapes, it can
never be recaptured. Here, however, the opposite is true: the two
dislocations can never truly escape from each other, due to the periodic
boundary conditions. At low temperatures, the mobile dislocation is
very likely to be ``recaptured'' and annihilated by its antiparallel
partner after completing a single revolution, as depicted in Fig.
\ref{escape_prob}. At finite temperatures, more revolutions are possible
before recapture.

\subsection {Nucleation of gliding dislocations along the cylinder axis}

Motivated by the bacterial elongation problem, in the previous example
the dislocations had a finite \textit{climb} mobility. For a crystal
of colloids, however, the motion of dislocations is predominately
via their \textit{glide}, as previously mentioned. With this motivation
in mind, we set $D_{c}=0$ in Eq. (\ref{eq:diffusion}) and consider
the unbinding of a pair of dislocation with $\vec{b}=\pm\vec{x}$
on a triangular lattice, where an external $\sigma_{xy}$ stress tensor
causes the dislocations to glide away from each other. Note that in
spite of the periodicity condition induced by the cylindrical geometry,
this component of the stress tensor is \textit{not} quantized, as
explained in Appendix B.

The $\sigma_{xy}$ stress creates a force along the $\hat{x}$ axis,
which is the glide direction. In this case, we have a \textit{one-dimensional}
escape-over-a-barrier problem, similar to that of Ref. \cite{bruinsma},
where the escape problem is solved in flat space. The potential in
our case is different due to the cylindrical geometry, and is given
by:
\begin{multline}
U(x)=\frac{Ab^{2}}{2}{\rm {log}}[{\rm \frac{2R}{b}sinh}(\pi(x-iy)/W)]+{\rm }\\
+\frac{Ab^{2}}{2}i\pi(x/W){\rm {csc}(\pi y/W){\rm {sinh}(\pi x/W){csch}(\pi(x-iy)/W)}}\\
+C.C-b\sigma_{xy}x,\label{energy-nucl}
\end{multline}
similar to the nucleation problem discussed above with $G=b\sigma_{xy}.$

For large enough $\sigma_{xy}$, the maximum of the potential occurs
for $x=Ab/\sigma_{xy}\ll W$, in which case the cylindrical geometry
plays no role. Let us consider the opposite limit, namely, $W\sigma_{xy}\ll Ab$.
In this case the maximum will occur for $x_{max}\gg W$, where the
potential decays exponentially. In infinite flat space, the barrier
for nucleation would diverge logarithmically as $\sigma_{xy}\rightarrow0$,
as discussed in Ref. \cite{bruinsma}. Here, on the other hand, the
interactions are bounded, and thus the nucleation barrier approaches
a \textit{constant $U_{max}$. }Upon using the one-dimensional version
of Eq. (\ref{langer}), we find that the prefactor of the Arrhenius
exponential term $e^{-U_{max}/T}$ is proportional to $\sqrt{U''(x_{max})},$
which vanishes as $\sigma_{xy}\rightarrow0$ and the maxima becomes
shallower. Using Eq. (\ref{energy_xx}) we find that that in this
case the prefactor vanishes as $\sqrt{{\sigma_{xy}}},$ up to logarithmic
corrections, so that:

\begin{equation}
\Gamma\propto\sqrt{{\sigma_{xy}}}e^{-U_{max}/T}.
\end{equation}

The above calculation represents a special case, since two antiparallel
dislocations with Burgers vectors making a generic, finite angle with
the $\hat{x}$ direction (say, $\theta$ and $\theta-\pi)$, will
have an interaction that provides a term \textit{linear} in $x$ for
$x\gg W$. Using Eq. (\ref{eq:yy_interaction_large_x}) and Eqs. (\ref{force_x-1})
we find that the component of the interaction force in the glide direction
is:

\begin{equation}
F_{int}=2\pi Ab^{2}\cos(\theta)\sin(\theta)/W
\end{equation}
A $\sigma_{xy}$ twist stress imposed on the boundaries of the cylinder
results in a force which has a component in the glide direction. Using
Eqs. (\ref{force_x}) and (\ref{force_y-1}) we find that:
\begin{equation}
F_{g}=F_{x}\cos(\theta)+F_{y}\sin(\theta)=b\sigma_{xy}[\cos^{2}(\theta)-\sin^{2}(\theta)].
\end{equation}

Thus, there will be a critical $\sigma_{xy}^{*}$ for which the interaction
term is compensated by the external stress, i.e., $F_{g}=F_{int}$:

\begin{equation}
\sigma_{xy}^{*}=\frac{2\pi Ab}{W}\tan(2\theta).
\end{equation}

At very low temperatures, the dislocations would no longer be able
to surmount the Peierls potential barriers. Then, the two dislocations
will glide away from each only for a much larger stress, $\sigma_{xy}\sim A$,
for which the force due to the external stress overcomes the periodic
Peierls potential \cite{hirth}. At finite temperature and $\sigma_{xy}>\sigma_{xy}^{*}$,
we obtain a similar escape-of-a-barrier problem we had in the previous
case, which can be solved in a similar manner. In this case, however,
the glide direction forms an angle $\theta$ with the $\hat{x}$-axis,
and once the dislocations unbind they make a helical trajectory along
to the glide direction.

\section{Acknowledgments}

We thank B. I. Halperin, W. T. M Irvine, F. Spaepen and V. Vitelli
for useful discussions. This work was supported by the National Science
Foundation via Grant DMR1005289 and through the Harvard Materials
Research Science and Engineering Laboratory, through Grant DMR0820484.
A.A. was supported by a Junior Fellowship of the Harvard Society of
Fellows.

\appendix
\section {The Airy stress function for a cylinder}

For some applications, it is convenient to describe the elastic stresses
in terms of the Airy stress function $\chi$, defined by $\sigma_{ij}(\vec{r})=\epsilon_{im}\epsilon_{jn}\partial_{m}\partial_{n}\chi(\vec{r})$
, which is a solution of the bi-harmonic equation \cite{hirth,LL}.
For a set of dislocations $\{\vec{b}_{\alpha}\}$ located at positions
$\{\vec{r}_{\alpha}\}$ with zero external stress, the Airy function
satisfies \cite{nelson_book}:

\begin{equation}
\nabla^{2}\chi(\vec{r})=4\pi A\sum_{\alpha}\epsilon_{ij}b_{\alpha i}\partial_{j}\delta(\vec{r}-\vec{r}_{\alpha}).
\end{equation}
We can obtain $\chi(x,y)$ from the strains, via: $\frac{\partial^{2}\chi^{x}}{\partial x^{2}}=\sigma_{yy}$,
$\frac{\partial^{2}\chi^{x}}{\partial x^{2}}=\sigma_{yy}$, $\frac{\partial^{2}\chi^{x}}{\partial x\partial y}=-\sigma_{xy}.$
Since for a dislocation with $\vec{b}=b\hat{x}$, $b\sigma_{xy}^{x}=\frac{\partial E_{x,-x}}{\partial x}=-b\frac{\partial^{2}\chi^{x}}{\partial x\partial y}$
and $-b\sigma_{xx}^{x}=\frac{\partial E_{\hat{x,}-\hat{x}}}{\partial y}=-b\frac{\partial^{2}\chi^{x}}{\partial^{2}y}$,
we find:

\begin{equation}
\frac{\partial\chi_{x}}{\partial y}=-(1/b)E_{\hat{x,}-\hat{x}}(x,y).
\end{equation}
 Upon integration this leads to:

\begin{eqnarray}
 &  & \chi^{x}(x,y)=\nonumber \\
 &  & \frac{Ab}{2W}\{\pi xy-xW{\rm {arctan}[{\rm {cot}(\pi y/W){\rm {tanh}(\pi x/W)]}}}\nonumber \\
 &  & +(ix+y){\rm W{log}\left[1-e^{2\pi(iy-x)/W}\right]}\}\nonumber \\
 &  & -\frac{Aby}{2}{(y+ix){\rm {log}[{\rm {sinh}[\pi(x-iy)/W]]}}}\label{airy_x}\\
 &  & -\frac{AbW}{4\pi}\left(i{\rm {PolyLog}\left[2,e^{-2\pi(x-iy)/W}\right]}\right)+C.C.\nonumber \\
\end{eqnarray}

Although one could add an undetermined function $G(x)$ to this expansion,
the derivatives of $\chi^{x}$ with $G(x)=0$ satisfy the above equations,
and thus Eq. (\ref{airy_x}) is the desired solution.

Similarly, for a dislocation with $\vec{b}=b\hat{y}$, we find:

\begin{equation}
\frac{\partial\chi_{y}}{\partial x}=(1/b)E_{\hat{y},-\hat{y}}(x,y),
\end{equation}
 which upon integration leads to:

\begin{eqnarray}
 &  & \chi^{y}(x,y)=\nonumber \\
 &  & \frac{-iAby}{2}\{\text{Log}\left[1-e^{-2\pi(x-iy)/W}\right]\nonumber \\
 &  & +\text{Log}(i\text{Sinh}[\pi(x+iy)/W])\}\nonumber \\
 &  & -\frac{Abe^{-i\pi y/W}(-i+\text{Tan}[\pi y/W])}{4\text{Sec}[\pi y/W]}\{y\text{Log}\left[1-e^{-2\pi(x-iy)/W}\right]\nonumber \\
\nonumber \\
 &  & (-ix+y)\text{Log}[\text{Sinh}[\pi(x+iy)/W]]\}+C.C.
\end{eqnarray}

Although adding an undetermined function$H(y)$ is again possible,
one can check that for $H(y)=0$ we have $\frac{\partial\chi^{y}}{\partial x^{2}}(x,y)=\sigma_{yy},$
$\frac{\partial\chi^{y}}{\partial y^{2}}(x,y)=\sigma_{xx},\frac{\partial\chi^{y}}{\partial x\partial y}(x,y)=-\sigma_{xy}$,
as required.

\section {Affine deformations of crystals on a cylinder}

Consider a lattice on a cylinder defined by minimal unit basis vectors
$\hat{e}_{1}$ and $\hat{e}_{2}$. As discussed in section \ref{phyllo},
the lattice can be characterized by a pair of integers $(M,N)$ such
that:

\begin{equation}
Mb\hat{e}_{1}+Nb\hat{e}_{2}=W\hat{y}.
\end{equation}
Upon neglecting uniform translations, a general affine deformation
of this lattice can be written as:

\begin{align}
u_{x}(x,y) & =\alpha_{xx}x+\alpha_{xy}y\nonumber \\
u_{y}(x,y) & =\alpha_{yx}x+\alpha_{yy}y.
\end{align}
These displacements will also result in a crystal, characterized by
different unit vectors $\hat{g}_{1}$ and $\hat{g}_{2}$. The actual
values of the matrix $\textbf{\ensuremath{\mathbf{{\bf \bm{{\alpha}}}}}}$
will be determined by boundary conditions, such as counter-rotating
twists applied to the ends of the cylinder. In general, the deformation
matrix will not be symmetrical, although (neglecting a weak dependence
of the lattice orientation relative to the cylinder axis), the energy
has to be independent of the antisymmetric part $\frac{1}{2}(\alpha_{xy}-\alpha_{yx})$.
From the periodicity in the $y$ direction we find that:
\begin{equation}
u_{x}(x,y+W)=u_{x}(x,y)+b(m\hat{g}_{1}+n\hat{g}_{2}),
\end{equation}
where $m$ and $n$ are integers, leading to:

\begin{equation}
\alpha_{xy}=\frac{b}{W}(m\hat{g}_{1,x}+n\hat{g}_{2,x}),\:\alpha_{yy}=\frac{b}{W}(m\hat{g}_{1,y}+n\hat{g}_{2,y}).\label{eq:quantize}
\end{equation}
Note that for small stresses corresponding to integers $m,n$ which
are not too large, we can replace the basis vectors $\hat{g}_{1}$
and $\hat{g}_{2}$ by $\hat{e}_{1}$ and $\hat{e}_{2}$, up to corrections
of order $(b/W)^{2},$ which is a higher order effect.

From Eq. (\ref{eq:quantize}) we conclude that the components of the
deformation tensor $\alpha_{xy}$ and $\alpha_{yy}$ are quantized
on a cylinder, while the components $\alpha_{xx}$ and $\alpha_{yx}$
are \textit{not} quantized. Considering the strain matrix $u_{ij}=\frac{1}{2}(\partial_{i}u_{j}+\partial_{j}u_{i})$,
we find that $u_{xx}=\alpha_{xx}$ is \textit{not} quantized, since
we have no restrictions on $u_{x}$, while $u_{yy}=\alpha_{yy}$ is
quantized. The off-diagonal component $u_{xy}=\frac{1}{2}(\alpha_{xy}+\alpha_{yx})$
is a combination of a quantized object and a non-quantized object,
and is thus \textit{non}-quantized. The stress tensor is related to
the strain tensor via the relation:
\begin{equation}
\sigma_{ij}=2\mu u_{ij}+\lambda\delta_{ij}u_{kk}.
\end{equation}
.

Each of the components of the stress tensor has a non-quantized piece.
Thus, the Peach-Koehler stress is fact non-quantized. Note, however,
that \textit{particular} linear combinations of the stress tensor
components can be quantized, similar to the case of superfluid films
on cylinders \cite{machta,nelson_amir_tbp}.

%\bibliographystyle{apsrev4-1}
%\bibliography{ariel_bib,bacteria}

\begin{thebibliography}{44}
\bibitem [{wee()}]{weertman_comment_PRX}%
  \BibitemOpen
  \href@noop {} {}\bibinfo {note} {J. Weertman (Phil. Mag. \textbf{11}, 1217
  (1965)) has suggested that only the \emph{traceless} part of the stress
  tensor should enter Eq. (8), a view which also appears in recent editions of
  Ref. [18]. Here, however, we take the more conventional perspective of Lothe
  and Hirth (J. App. Phys. \textbf{38}, 845 (1967)), which we believe is the
  appropriate one for cylindrical geometries where new material is supplied
  from the bulk for the climb process.}\BibitemShut {Stop}%
\end{thebibliography}
%merlin.mbs apsrev4-1.bst 2010-07-25 4.21a (PWD, AO, DPC) hacked
%Control: key (0)
%Control: author (72) initials jnrlst
%Control: editor formatted (1) identically to author
%Control: production of article title (-1) disabled
%Control: page (0) single
%Control: year (1) truncated
%Control: production of eprint (0) enabled
%

\end{document}